\documentstyle[11pt,aaspp4,flushrt]{article}

\def\Lya{Ly$\alpha\ $}
\def\Lyb{Ly$\beta\ $}
\def\HI{\hbox{H~$\rm \scriptstyle I\ $}}
\def\HII{\hbox{H~$\rm \scriptstyle II\ $}}
\def\DI{\hbox{D~$\rm \scriptstyle I\ $}}
\def\HeI{\hbox{He~$\rm \scriptstyle I\ $}}
\def\HeII{\hbox{He~$\rm \scriptstyle II\ $}}
\def\HeIII{\hbox{He~$\rm \scriptstyle III\ $}}
\def\CIV{\hbox{C~$\rm \scriptstyle IV\ $}}
\def\NHI{N_{\rm HI}}
\def\NHeII{N_{\rm HeII}}
\def\cm2{\,{\rm cm$^{-2}$}\,}
\def\kms{\,{\rm km\,s$^{-1}$}\,}
\def\ev{\,{\rm eV\ }}
\def\kel{\,{\rm K\ }}
\def\intunits{\,{\rm ergs\,s^{-1}\,cm^{-2}\,Hz^{-1}\,sr^{-1}}}
\def\ltsima{$\; \buildrel < \over \sim \;$}
\def\lsim{\lower.5ex\hbox{\ltsima}}
\def\gtsima{$\; \buildrel > \over \sim \;$}
\def\gsim{\lower.5ex\hbox{\gtsima}}
\def\etal{{ et~al.~}}

\def\mur86{Murdoch \etal 1986}
\def\hu95{Hu \etal 1995}
\def\lu91{Lu \etal 1991}
\def\luu96{Lu \etal 1997}
\def\gia91{Giallongo 1991}
\def\bec94{Bechtold 1994}
\def\bah93{Bahcall \etal 1993}
\def\bahh96{Bahcall \etal 1996}

\begin{document}

\title{Spectral Analysis of the Lyman Alpha Forest
       in a Cold Dark Matter Cosmology}

\author{Yu Zhang$^{1,2}$,
            Peter Anninos$^{1}$,
            Michael L. Norman$^{1,2}$, and
            Avery Meiksin$^{3,4}$}

\vskip10pt
\affil{\em $^1$Laboratory for Computational Astrophysics, National Center
for Supercomputing Applications, University of Illinois at Urbana-Champaign,
405 N. Mathews Ave., Urbana, IL\ 61801}

\vskip10pt
\affil{\em $^2$Astronomy Department, University of Illinois at Urbana-
Champaign, 1002 West Green Street, Urbana, IL \ 61801}

\vskip10pt
\affil{\em $^3$Department of Astronomy \& Astrophysics, University of
Chicago, 5640 South Ellis Avenue, Chicago, IL\ 60637}

\vskip10pt
\affil{\em $^4$Edwin P. Hubble Research Scientist}

\date{\today}

\begin{abstract} We simulate the \Lya forest in a standard CDM
universe using a 2--level hierarchical grid code to evolve the dark
and baryonic matter components self--consistently. We solve the
time--dependent ionization equations for hydrogen and helium, adopting
the Haardt \& Madau (1996) estimate for the metagalactic UV radiation
background. We compare our simulation results with the measured
properties of the \Lya forest by constructing synthetic spectra and
analyzing them using an automated procedure to identify, deblend, and
fit Voigt line profiles to the absorption features.
The \HI column density and Doppler--parameter
distributions we obtain agree closely with those measured by the
Keck HIRES and earlier high spectral resolution observations
over the column density range $10^{12}$ \cm2 $< \NHI <
10^{16}$ \cm2. In particular, we find a power--law column density
distribution persists to the lowest values ($\sim10^{12}$ \cm2)
reported from the HIRES measurements, in agreement with the
incompleteness--corrected observations. We are able to match the normalization
of the column density distribution at the low end using the Haardt \& Madau
spectrum and a baryon density consistent with nucleosynthesis limits. We find,
however, a significant deficit of systems at higher column densities,
$\NHI>{\rm few}\times10^{16}$ \cm2. The deficit arises from a curvature in the
column density distribution that may not be removed by an overall
renormalization. We find evolution in the cloud
number density and opacity comparable to the observed evolution,
though growing somewhat too quickly at the highest redshifts. The
evolution, however, is sensitive to the assumed UV radiation field,
which becomes increasingly uncertain at higher redshifts ($z>3.5$).
We also compare with measured values of the intergalactic \HeII opacity.
Our results require a \HeII ionizing background lower than the Haardt \& Madau
estimate by a factor of 4, corresponding to a soft intrinsic QSO spectrum
of $\alpha_Q\approx1.8-2$.

\end{abstract}


\newpage

\section{Introduction}
\label{int}

Since the identification of \Lya absorption lines in QSO spectra
(Lynds 1971), and the first major survey of their properties (Sargent
\etal 1980), the \Lya forest has been recognized as a unique probe of the
physical state of the universe during the early stages of cosmic
evolution.  Studies of these absorption features offer several
advantages over galaxies: (1)\ the lines are associated
with typical structures as opposed to the ``rare'' structures (such
as galaxies and clusters) which arise from the statistical tail of
density perturbations; (2)\ they are observed along random directions
in the sky and may not be biased with respect to the local structure
distribution; (3)\ they are found at very high redshifts and thus
probe the matter content at early times when the differences between
competing cosmological models are most pronounced; and (4)\ they
potentially comprise an enormous database with approximately $10^3$
discovered QSOs with each QSO generating tens to hundreds of lines,
presenting a distinctive statistical advantage compared to other astronomical
observations.  Using remote high--redshift QSOs as beacons through the
intergalactic medium (IGM), the universe can be probed on much larger
scales than allowed by existing catalogs of galaxies.  They can be
used both as tracers of the matter distribution and as probes of
physical conditions in the universe at very early epochs when larger
structures, such as galaxies, are still forming and evolving.  \Lya
clouds may be the missing link between primordial density fluctuations
and the formation of galaxies and clusters of galaxies.

Early attempts to explain the \Lya forest lines attributed the
absorption features to the hydrogen in intervening protogalaxies
photoionized by a QSO-dominated UV background (Arons 1972), or to
self--gravitating isothermal gas clouds (Black 1981), although the
latter are gravitationally unstable over the range of column densities
required. Bahcall \& Salpeter (1965) argued that broad \Lya absorption
features should appear due to intervening clusters of galaxies, though
these have yet to be definitively detected.  Ostriker \& Ikeuchi
(1983) (see also Ikeuchi \& Ostriker 1986), following the lead of
Sargent \etal (1980), subsequently formulated an elaborate theory of
intergalactic clouds stabilized by the pressure of an external hot
IGM.  The virtue of this theory is that it provides several
observational tests and predictions about both the IGM and the clouds
themselves. However, the theory lacks a clear picture for the
formation of \Lya clouds and ionization of the IGM.  Shortly
thereafter, the successes of the CDM model created an interest in
combining this theory of large scale structure formation with the
formation of \Lya clouds into a single unified picture.  In this
scenario, \Lya clouds can either be associated with bound structures,
stabilized by the gravity of dark matter mini--halos (Rees 1986, 1988;
Ikeuchi 1986), or with post-photoionized unconfined gas in low mass
objects that developed from small mass fluctuations (Bond, Szalay, and
Silk 1988).

These theoretical models, however, treat \Lya clouds as isolated objects and
attempt only to explain their physical state and
interactions with the environment.
They do not provide a self--consistent framework for the formation,
evolution, and interactions of the clouds, nor can any single analytical model
account for the broad range of observed cloud properties.
There are also inherent limitations to these models because
of their underlying assumptions and simplifications; for example, neglecting
or simplifying microphysical processes and the nonlinear interactions
of the small scale fluctuations which collapse at nearly the same time
in CDM--like models.

The limited ability of analytic theories to account for the complexity
and range of the observed properties of \Lya clouds and the IGM, combined
with the rapid progress in numerical techniques, have motivated a number of
attempts to study the formation and evolution of the intergalactic
medium as a whole, in both clumpy and diffuse forms, by means
of cosmological hydrodynamics/$N$--body simulations. The advantage of this
approach is that it is not necessary to impose {\it a priori} any
physical constraints on the clouds (i.e., pressure or gravity confinement)
nor on their environments. Given a spectrum of primordial perturbations,
the clouds form and evolve naturally
under gravitational, hydrodynamic, thermal and chemical forces.

Three major cosmological simulations of QSO absorption line systems
have been performed to date: The first \Lya simulation was performed
by Cen \etal (1994) (see also Miralda--Escud\'e \etal 1996)
in a flat CDM universe with a nonzero cosmological
constant using a grid--based TVD hydrodynamics code. Zhang, Anninos, \& Norman
(1995), hereafter referred to as ZAN95, performed a
self--consistent multi--species simulation in a CDM
universe using a 2--level hierarchical grid code for higher resolution.
Independently, Hernquist \etal (1996) performed a
simulation of the standard CDM (SCDM) universe, but using
a Smooth Particle Hydrodynamics (SPH) code. Significantly, all the
calculations are able to recover the general statistical properties
of the \Lya clouds, such as the column density and equivalent
width distributions, Doppler parameters, and the size of the
absorption features (Charlton \etal 1996).
A theoretical paradigm is thus beginning to
emerge from these calculations in which \Lya absorption lines originate
from the small scale structure in pregalactic or intergalactic gas through
the bottom--up hierarchical formation picture in a CDM--like universe.
The absorption features originate in structures exhibiting a variety of
morphologies, from fluctuations in underdense regions to spheroidal minihalos
to filaments extending over scales of a few megaparsecs. A more detailed
discussion of the morphologies and physical state of the systems found in
our simulations is presented in Zhang \etal (1997).

In this paper, we follow up on our previous effort (ZAN95) to
present a unified picture of the IGM and \Lya absorbers for a CDM universe.
The results in ZAN95 are extended in
\S \ref{sec:stat_sta} and \S \ref{sec:stat_evo} to
include studies of the line center opacity, metal line systems,
flux decrement calculations, and evolutions of the effective opacity
and line number,
in addition to presenting more detailed discussions of
the Doppler parameters and column density and
equivalent width distributions at a fixed redshift $z=3$.
A comprehensive discussion of the statistical correlation between
the observed spectral properties of \Lya lines and the physical state of
the absorbers will be presented in a companion paper (Zhang \etal 1997).

Furthermore, we note that the statistical analysis
presented in ZAN95 was derived from data obtained directly from the physical
properties of the baryonic
gas in the simulation. Here we describe a new procedure
we developed to more closely mimic observational methods.
In particular, we have implemented a convolution
algorithm to synthesize absorption spectra through the computational box, and
developed a new analysis procedure to identify, deblend, fit and extract
the physical properties of absorption features in generated spectra.
It should be pointed out that Cen \etal (1994)
and Hernquist \etal (1996) also utilize a spectral method to synthesize
and reduce spectra. However, they adopt a simplified ``threshold'' method
and do not attempt to deblend lines in analyzing their spectral data.
The failure to properly deblend lines will result in an overestimate of
the Doppler widths of the lines. Their technique will also substantially
undercount the lowest column density features in the spectra.
More advanced methods of analysing spectral data, such as those introduced
here, are required to better assess the agreement with the observed data.
A similar approach has recently been adopted by Dav\'e et al. (1997).
Following the discussion of our numerical methods and simulations in
\S \ref{sec:simulations}, we present a detailed investigation
of the different spectral analysis methods applied to our
numerical data in \S \ref{sec:spec}. A detailed analysis of the derived
statistics of the \Lya forest at $z=3$ is provided in \S \ref{sec:stat_sta},
and of the evolution of the forest in \S \ref{sec:stat_evo}. We summarize
our results in \S \ref{sec:stat_sum}.

\section{The Simulations}
\label{sec:simulations}

Altogether, we have performed four separate simulations of the \Lya
forest in a standard Cold Dark Matter (CDM) dominated universe using
our 2--level hierarchical grid code HERCULES (Anninos, Norman, \&
Clarke 1994; Anninos \etal 1996).  The different physical and
computational parameters for each of the calculations are summarized
in Table \ref{tab:sim_res}.  Two different box sizes, each with two
levels of nested grids, were considered: 3.2 and 9.6 Mpc, a reasonable
compromise between resolving small scale structures, including larger
scale power, and sampling a large enough volume to derive reliable
statistics.  The smaller 3.2 Mpc box simulations were discussed in
ZAN95. Here we focus on the results from the larger 9.6 Mpc box, and
merely reference and make passing comparisons to the smaller box
results when appropriate.

Our model background spacetime is a flat, cold dark matter dominated universe
with the initial density perturbations originating from inflation--inspired
adiabatic fluctuations.  The BBKS (Bardeen \etal 1986) transfer function
is employed with the standard Harrison--Zel'dovich power spectrum.
We adopt a fluctuation amplitude consistent with the present number density
and temperatures of galaxy clusters (White, Efstathiou, \& Frenk 1993;
Bond \& Myers 1996), although we also compare with the
earlier high amplitude model. While the CDM power-spectrum is known not
to be able to match both the clustering of galaxies at small scales and the
measurements of COBE at large, it adequately represents the power-spectrum
over the range of scales of interest to the \Lya forest problem.
The baryonic fraction $\Omega_b$ is chosen so that $\Omega_bh^2$ is
consistent with the latest results from Big Bang nucleosynthesis
(Copi, Schramm, \& Turner 1995), and the
baryonic fluid is composed of hydrogen and helium in primordial abundance
with a hydrogen mass fraction of 76\%.
We use the COSmological initial conditions and MICrowave anisotropy codeS
(COSMICS, Bertschinger 1995)
to generate the initial particle positions and velocity perturbations
appropriate to the particular parameters in the large box simulations.
The cosmological model parameters used in our different simulations
are provided in Table \ref{tab:sim_res}.

For the most part, the statistical analysis of the \Lya forest
is performed with the top grid evolution as it contains a
greater sample or range of data. However,
to explore the effects of grid resolution, we take advantage of the
2--level nature of our code and introduce a second, more finely resolved, grid
to cover a sub--region of the coarser top grid.
In all our simulations, we use $128^3$ cells on both the top and subgrids;
$64^3$ dark matter particles were used in the small 3.2 Mpc box simulation.
However, the larger box calculation included $128^3$ dark matter particles
to verify that the particle sampling does not significantly affect the
fluctuations in the cosmic voids.
The subgrid is centered on the {\it least} dense region of the top grid
for the purpose of resolving in greater detail the lowest column density
absorbers which are found in the voids (Zhang \etal 1997).
The refinement factor (the ratio of cell sizes
between the top and sub grids) is 4 for all the subgrid simulations,
yielding an effective grid resolution of $512^3$ cells.
Table \ref{tab:sim_res} also lists the spatial and mass resolutions for
both the top and subgrids of the different simulations.
The characteristic length scale of the \Lya forest clouds, both observed
and simulated, is typically
a hundred or more kiloparsecs (Smette \etal 1995; Dinshaw \etal 1995;
Fang \etal 1996; Charlton \etal 1996; Hernquist \etal 1996),
and it is generally believed that their typical mass
is of order $10^7 - 10^9~M_\odot$.
Table \ref{tab:sim_res} clearly demonstrates
that our numerical setup has more than sufficient spatial and
mass resolution to probe scales smaller than the typical \Lya clouds.

In addition to the usual ingredients of baryonic and dark matter,
we also solve the coupled system of non--equilibrium chemical reactions with
radiative cooling. The reaction network includes a
self--consistent treatment of the following six species:
\HI, \HII, \HeI, \HeII, \HeIII
and $e^-$. Details of the
chemical model, cooling rates and numerical methods can be found in
Abel \etal (1996) and Anninos \etal (1996). A final but essential component
introduced in our simulations is an
ultraviolet (UV) radiation background to ionize the otherwise
neutral intergalactic medium.  The intergalactic clouds are optically thin
at the Lyman edge for column densities smaller than about $\NHI = 8 \times
10^{16}$ \cm2 (corresponding to $\tau_L\approx0.5$, where $\tau_L$ is the
\HI opacity at the Lyman edge).
Hence, for the range of column densities we are interested in,
the optically thin limit is a good approximation,
i.e. it is not necessary to account for self-shielding of the external
ionizing radiation field.  The radiation background is also assumed
to be uniform which is a fairly good approximation if one is
mainly interested in the statistical properties of the \Lya absorbers.

A simple though somewhat arbitrary parameterization of the UV background
radiation field $J_\nu$ is a power law: $J_\nu=J_L(\nu/\nu_L)^{-\alpha}$,
where $\nu_L$ is the \HI frequency at the Lyman edge and the
spectral index $\alpha$ is in the range 1 to 2.
The amplitude of the radiation intensity is measured to be
$10^{-22\pm1}\intunits$, according to the QSO proximity
effect (Bajtlik, Duncan, \& Ostriker 1988; Lu \etal 1997).
The redshift dependence of the amplitude and the epoch at which the radiation
is turned on must also be specified.
Such a parametric prescription was utilized in our ZAN95 simulations.

\begin{figure}
\plottwo{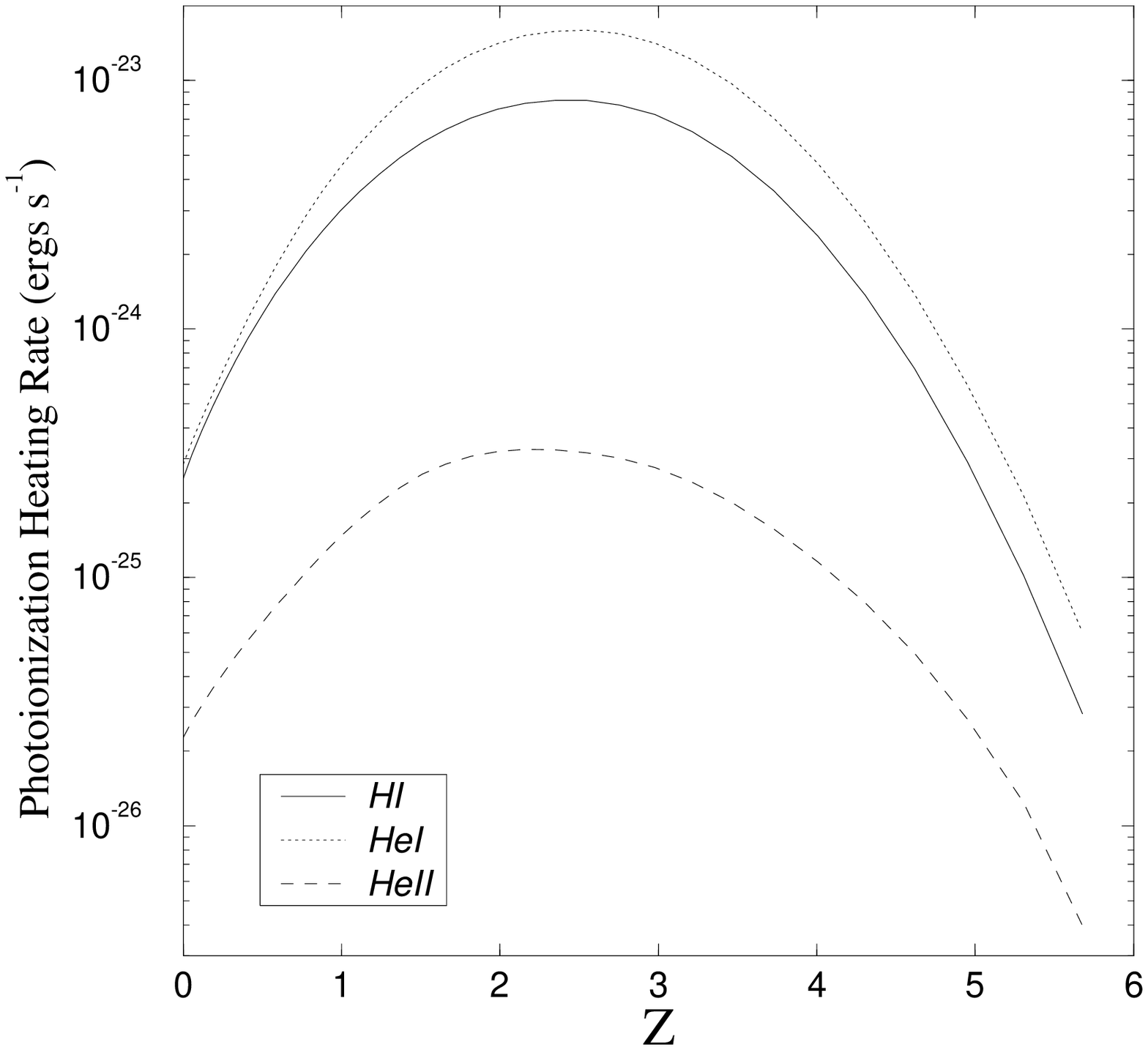}{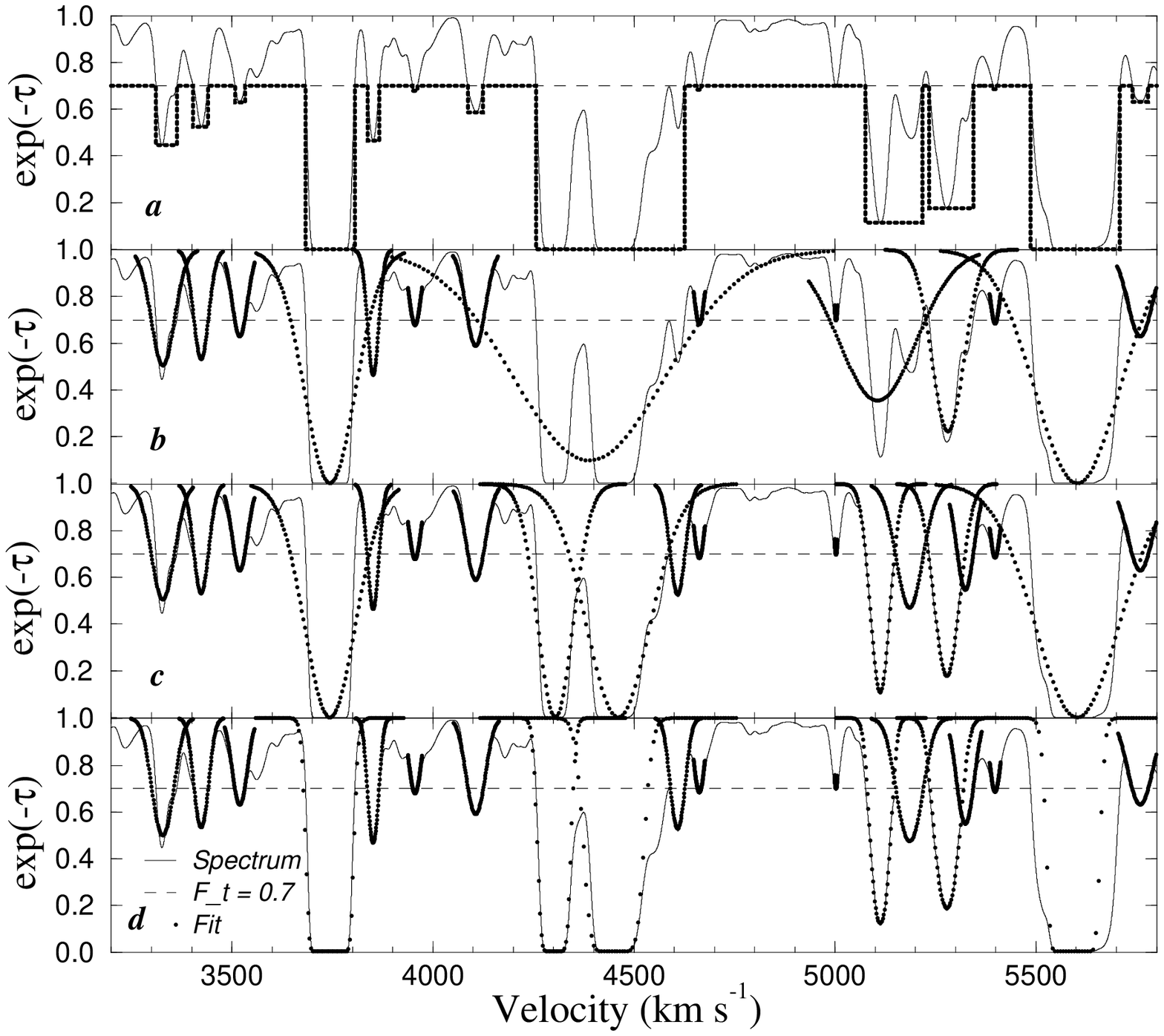}
\caption{Photoionization heating rates per atom as a function of redshift
for \HI, \HeI, and \HeII from the Haardt and Madau (1996) radiation field.}
\label{fig:HMradiation}
\caption{Comparison of the different spectral line analysis methods:
(a) the threshold method without profile fitting;
(b) the threshold method with profile fitting;
(c) the deblending method applied to the spectral data; and
(d) the deblending method applied to the opacity data.
The solid lines are the actual spectra, generated at $z=3$,
and the dots represent the identified lines and their fitted profiles.
The horizontal dashed lines indicate the spectral cutoff used
in this example, $F_t=0.7$.}
\label{fig:spec_com}
\end{figure}

Haardt \& Madau (1996), however, have recently computed the
UV radiation field with radiation transfer
in a clumpy universe, based on the measured luminosity function of QSOs. The
field is found as the self-consistent solution of the combined system of QSO
sources, absorption by the \Lya forest and Lyman limit systems, and reemission
of the recombination radiation from the absorbing clouds. The photoionization
rates $\Gamma$ are provided by Haardt and Madau. We use their spectra
to generate the photoionization heating rates, and fit them to the same
functional form as Haardt and Madau adopt for the photoionization rates.
These are shown plotted in Figure \ref{fig:HMradiation}.
We use the fits for the photoionization and photoionization heating rates
in this paper. Since the cloud temperatures are
not very sensitive to the assumed heating rate, we note that comparisons
of our current results with those in ZAN95 can be made by simply
rescaling by the ratio of radiation intensities, or more precisely,
by the ratio of parameters $\Omega_b^2/\Gamma$, where $\Gamma$ is
the photoionization rate for the species of interest
(see \S \ref{subsec:colden}). The temperature is somewhat dependent on the
gas density (Meiksin 1994; Croft \etal 1997), but for a limited range in
$\Omega_b$ (a factor of about 2), this is an approximately valid procedure,
provided the regions have not undergone significant recombination into neutral
hydrogen or neutral or singly ionized helium.

\section{Spectral Analysis Methods and Tests}
\label{sec:spec}

To compare the simulation results with the measured properties of the \Lya
forest, it is desirable to mimic the observed data and its analysis
as closely as possible. To do so, however, would entail building a
sophisticated spectral simulation procedure that includes the vagaries
of the actual data. This would require modeling the noise in the
sky--subtracted spectra, the variations in the QSO spectrum through the forest,
continuum fitting, and accounting
for the finite resolution of the detectors. Rather than embarking on a
massive simulation program at this stage, we instead concentrate on extracting
the actual underlying statistical distributions of the absorber spectral
properties that would be found from ``perfect'' observations. The belief is
that the two distributions should coincide for optically thick (at line center)
systems that are not too badly saturated, though they may deviate for the
weaker
optically thin systems. For the latter, we rely on the Monte Carlo simulations
performed by the observers to test their success in recovering the weakest
lines to evaluate the ability of the simulations to match the observational
data. In this section, we describe the methods we have developed to generate
and analyze our spectra. We also compare the results of our analysis method,
which is based on Voigt profiles, against the
threshold approach, as the latter was adopted in the most comprehensive
treatments of the properties of the forest in previous simulations.

\subsection{Generating Spectra from Numerical Data}
\label{sec:spec_gen}

\Lya absorption in the spectrum of a QSO located at redshift $z$
is calculated by $e^{-\tau_\nu}$, where the optical depth $\tau_\nu$ is given
by

\begin{equation}
\label{eqn:spec_taut}
\tau_\nu(t) = \int_t^{t_0} n_{\rm HI}(t)~\sigma_\nu~c~dt~,
\end{equation}
and the integration is performed along the line of sight between the QSO and
observer.  Here $c$ is the speed of light, $n_{\rm HI}$ the number density of
the \HI absorbers, and $\sigma_\nu$ the absorption cross-section which
can be expressed as a Voigt function (Spitzer 1978),
\begin{equation}
\label{eqn:spec_sig}
\sigma_\nu = \sigma_0~V \approx \frac{\sigma_0}{\sqrt{\pi}\Delta\nu_{\rm D}}
             \exp\left[-\frac{(\nu-\nu_0)^2}{(\Delta\nu_{\rm D})^2}\right]
             + \frac{\sigma_0\Delta{\nu_{\rm Lor}}}{2\pi(\nu-\nu_0)^2},
\label{sigmanu}
\end{equation}
where $\nu_0$ is the \Lya rest frequency,
$\sigma_0=({\pi}e^2/m_ec)f$ is the resonant \Lya cross section
and $f$ is the upward oscillator strength.

In the first term of equation (\ref{sigmanu}), $\Delta\nu_{\rm D}$ is
proportional to the full width half maximum of the Gaussian profile
due to the Doppler broadening which dominates the profile core
\begin{equation}
\label{eqn:spec_nuD}
\Delta\nu_{\rm D} = \nu_0\frac{b}{c}
            = \nu_0\sqrt{\frac{2kT}{{m_{\rm p}}c^2}}~,
\end{equation}
where $T$ is the gas temperature, $k$ Boltzmann's constant
and $m_{\rm p}$ the proton mass.
The second term in equation (\ref{sigmanu})
accounts for the asymptotic behavior of the
Lorentzian wings of the profile as a result of radiation damping.
The full width half maximum of the Lorentzian profile is given by
\begin{equation}
\label{eqn:spec_nuL}
\Delta\nu_{\rm Lor} = \frac{4{\pi}e^2}{3{m_e}c^3}{\nu_0}^2f
              \approx 4\times10^{-23}{\nu_0}^2f\,{\rm Hz}~.
\end{equation}
Since the Lorentzian wing is negligible for lines
with equivalent width $W_0\leq1$~\AA~ and
column density $\NHI\leq10^{19}$ \cm2, we may use the Doppler term alone
for the low and medium column density \HI
absorption lines. The radiation damping wings, however,
can be significant for the \HeII lines, so a full Voigt profile
is used in these cases.

Equation (\ref{eqn:spec_taut}) can then be parameterized, to order $v/c$,
by the redshifted frequency $\nu$ as (see also Bi, Ge, \& Fang 1995)
\begin{equation}
\label{eqn:spec_tauz}
\tau_\nu(z) = \frac{c^2 \sigma_0}{\sqrt{\pi} \nu_0}
              \int_z^{z_0} \frac{n_{\rm HI}(z')}{b} \frac{a^2}{\dot a}
              \exp\left\{-\left[(1+z')\frac{\nu}{\nu_0}-1+\frac{v}{c}\right]^2
                         \frac{c^2}{b^2}\right\} ~ dz'~.
\end{equation}
Here, $v$ is the peculiar velocity (coherent and turbulent)
along the line--of--sight (LOS) which produces a shift at the line center
relative to the Hubble velocity, and contributes to the line broadening
together with the thermal motion.  The cosmological expansion rate
in equation (\ref{eqn:spec_tauz}) is given by the Friedman equation
for the cosmological scale factor $a$,
\begin{equation}
\label{eqn:spec_adot}
\dot a = H_0 \sqrt{1+\Omega_0 z + \Omega_\Lambda \left(a^2 -1 \right)}~.
\end{equation}

To synthesize an absorption spectrum over a large redshift interval, say from
$z=3$ to $z=0$, we divide this redshift range into equal intervals of
$\Delta z=0.1$ (the smallest practical interval at which we have retained
data from the 3D simulations).  A LOS is chosen randomly and the
integrated optical depth (and absorption spectrum) is computed from
equation (\ref{eqn:spec_tauz})
for different values of redshifted frequency $\nu$ in each redshift interval.
To resolve the smallest spectral features present in our simulations, we use
50000 points to cover the range $\Delta z=0.1$. The redshift interval
corresponding to a physical length scale at any redshift $z$ is given by
\begin{equation}
\label{eqn:spec_dz}
\delta z = 3.34 \times 10^{-4} \frac{L h}{a}
           \sqrt{\Omega_0 z + \Omega_\Lambda (a^2 -1) + 1}~,
\end{equation}
where $L$ is the comoving box size in units of Mpc.  For example,
at redshift $z=3$ in an $\Omega_0=1$ universe, equation
(\ref{eqn:spec_dz}) indicates
that a box size of 9.6 Mpc corresponds to a redshift interval of
$\delta z = 1.3\times10^{-2}$. Hence, each LOS samples the structures
in the same cube approximately eight times over the interval
$\Delta z = 0.1$.
However, choosing a random
orientation vector assures that each LOS samples the cube uniformly
and does not bias the synthesized spectra in any significant way.

Table \ref{tab:spec_res} lists the redshift, velocity and wavelength
resolutions
of our numerically generated spectrum, along with the corresponding
resolutions of real observation instruments.
Our spectral resolution is quoted in redshift space
and is constant for all redshifts: $\Delta z=\Delta
v/c=0.1/50000=2\times10^{-6}$.
This resolution is more than adequate to cover the grid cell intervals,
which, at $z=3$, are
$\Delta z = 1\times 10^{-4}$ and $2.5\times 10^{-5}$ for the top and sub
grids respectively.
Keck/HIRES is the High Resolution Echelle Spectrograph installed on the
10-meter Keck telescope, and KPNO/KPE is the Kitt Peak Echelle spectrograph
on the 4-meter Mayall telescope at the Kitt Peak National Observatory.
These two spectrographs have obtained the bulk of the high resolution \HI
\Lya data to which we will be comparing our numerical results.
ASTRO-2/HUT is the Hopkins Ultraviolet Telescope aboard the ASTRO-2 shuttle
mission launched in 1994.  HST/FOC is the Faint Object Camera aboard the
Hubble Space Telescope launched in 1990.  These two instruments have
detected intergalactic \HeII absorption along the lines-of-sight to two QSOs
(Davidsen, Kriss, \& Zheng 1996; Jakobsen \etal 1994).

\subsection{Line Analysis}
\label{sec:spec_red}

Finding and fitting the absorption lines requires assumptions to be made
regarding the shapes of the line profiles. Ideally, one would like to
characterize the physical properties of the clouds imposing as little
prejudice as possible. In order to avoid imposing a specific profile
unnecessarily, Miralda-Escud\'e developed a ``threshold''-based method of
spectral analysis (Hernquist \etal 1996; Miralda-Escud\'e \etal 1996).
This method identifies lines as regions in the spectrum $e^{-\tau_\lambda}$
below a certain transmission threshold and computes the properties
of these lines, such as the equivalent width and column density,
using only the data below the threshold.
Specifically, they evaluate the following integrals for the equivalent width
and column density, respectively
\begin{equation}
\label{eqn:spec_wid}
W_0 = \int_{\lambda_1}^{\lambda_2} \left( 1 - e^{-\tau_\lambda} \right)
      \frac{d\lambda}{1+z}~,
\end{equation}
\begin{equation}
\label{eqn:spec_int}
\NHI = \int_{\lambda_1}^{\lambda_2} \frac{(1+z)c{\tau_\lambda}d\lambda}
         {{\sigma_0}{\lambda}^2}~,
\end{equation}
where $\lambda_1$ is the down--crossing wavelength where the transmission drops
below the threshold, and $\lambda_2$ is the up--crossing wavelength where the
transmission rises above the threshold with $\lambda_2>\lambda_1$.
This method has the advantage of simplicity, permitting many lines
to be identified and evaluated efficiently.
It also has the merit of not assuming any particular
line profile (except to compute the Doppler parameter $b$,
a Voigt profile must effectively be adopted).
For small line densities (at low redshifts) and with sufficiently
low transmission thresholds, this method is an effective means for
measuring the line properties.

For narrow and weak lines which are not blends, however, $W_0$ and $\NHI$ can
both be underestimated if too low a spectral threshold is chosen,
since the contribution from above the cutoff is not counted in the integrals.
In addition, since no deblending is attempted to help isolate distinct
components, $W_0$, $\NHI$ and $b$ can be overestimated
in the case of broad lines that are mostly blends. In fact,
$W_0$, $\NHI$ and $b$ are to first approximation the sums of the individual
components making up the line complex.
The first problem may be alleviated by setting
a high transmission threshold.  But this comes at the expense of introducing
more blended lines, exacerbating the second problem.  Table \ref{tab:spec_com}
(discussed in detail below) shows that overestimates of the line widths
are much greater than underestimates associated with too low a threshold.
Attempting to use a very low value for the threshold to avoid too many blends,
however, would result in missing most of
the narrow and low column density lines.

With the arrival of high resolution observational data and
a larger database of lines at higher redshifts, it is essential
to develop an alternative spectral reduction algorithm
that is better suited to dealing with strong blending as well as being
able to identify most of the individual absorption features in
the spectrum for high transmission cutoffs.
A line deblending procedure is essential to
extract meaningful statistics and to make a full
comparison with the new high resolution observational spectra.
While a comparison with integrated absorption properties of the
spectra may be made without line--identification, the lines contain
a wealth of detailed information concerning the structure of the spectra.
A complete comparison should include the effect of systematics and errors
as closely as possible to those of the observed spectra. At present this
is beyond the scope of the paper, though we anticipate making such a
comparison in the future. Our present attempt is to analyze the spectra
under ideal observing conditions. Indeed, it is the ideal results that
the observer attempts to extract from his or her spectra, often including
Monte Carlo simulations to correct for systematics (e.g., Hu \etal 1995).

We have developed a new procedure to identify the
absorption lines above a specified opacity threshold (or below a spectral
threshold) and to
deblend the identified absorption features into individual and isolated lines
using the sign and value of the local derivatives of the opacity (or
transmission)
with respect to the wavelength.  The use of extrema allows one to identify
local features within larger ones without relying on the spectrum to cross the
transmission threshold.  While the method is not foolproof --- it may still
miss weak lines superimposed on the wings of
much stronger ones --- it represents a
significant improvement over the simple threshold method,
especially for the lower column density systems.

After an individual line is identified and isolated, the rest-frame equivalent
width can be computed using equation (\ref{eqn:spec_wid}), performing
the integration over the full line profile and not just over the
region beneath the threshold.  In order to extrapolate the data above the
threshold and up to the continuum, a line profile needs to be assumed.
The $W_0$ thus computed has the advantage of being directly comparable
to observations, but at the price of assuming a particular shape for the
line profile. For the high transmission cutoff that we adopt ($F_t=0.95$), the
contribution to $W_0$ from above the threshold is negligibly small
(because of the exponential drop in the Gaussian tail) and the
integration over the region under the threshold is a good approximation to
the true equivalent width provided the feature under the threshold
is deblended.

To facilitate a comparison with the measured line properties, we
follow the observers in adopting a Voigt profile. For lines with \HI column
densities smaller than $10^{19}$ \cm2, the radiation damping wings contribute
negligibly to the profile. Accordingly, we simplify the analysis by adopting
a Doppler profile. The Doppler parameter $b$ of each line can be obtained in
one of two ways:\ either by fitting the isolated spectral (or opacity) profile
with a Doppler function and deriving $b$ from the line width,
or using the following formula relating $b$ to $W_0$ (Spitzer 1978),
\begin{equation}
\label{eqn:spec_b}
b = \frac{W_0 c}{2\lambda_0 F(\tau_0)}~,
\end{equation}
where $\lambda_0$ is the \HI \Lya rest wavelength,
$\tau_0$ is the optical depth at line center, and
\begin{equation}
\label{eqn:spec_Ftau}
F(\tau_0) = \int_0^\infty \left[1 - \exp\left(-\tau_0 e^{-x^2}\right)\right] ~
dx
\end{equation}
is the curve of growth function for a Maxwellian velocity distribution,
which can be found tabulated in Spitzer (1978).
For unsaturated lines $F(\tau_0)\propto\tau_0$, and
the equivalent width is on the linear part of the curve of growth,
i.e. $W_0\propto \NHI$, independent of $b$ (since $\tau_0\sim \NHI/b$).
For saturated lines $F(\tau_0)\propto(\ln\tau_0)^{1/2}$, and the equivalent
width is on the flat part of the curve of growth,
i.e. $W_0\propto b(\ln\tau_0)^{1/2}$, increasing slowly with $\NHI$.
We find that the two determinations of the Doppler parameter typically
agree to within a percent. We employ the first method for our data analysis.
The column density is then determined from the line center opacity
and Doppler parameter using (Spitzer 1978),
\begin{equation}
\label{eqn:spec_col}
\NHI = \frac{\sqrt{\pi} \tau_0 b}{\sigma_0 \lambda_0}~.
\end{equation}

We note that either the opacity $\tau_\lambda$ or the transmission
$e^{-\tau_\lambda}$ can be used in the spectral reduction procedure.  We have
tried both and found the differences to be negligible
for unsaturated lines.  However, for strongly
saturated lines ($\NHI\gsim10^{14}$ \cm2), the opacity data is needed
to obtain a good fit to the individual lines because the large opacity causes
the transmission to be vanishingly small and to resemble a square well,
making it more difficult to fit properly
(compare Figures \ref{fig:spec_com}c and \ref{fig:spec_com}d). In this regard,
we differ from Dav\'e \etal (1997) who confine their spectral analysis to
the transmitted flux. Dav\'e \etal (1997) instead adopt an approach that
more closely mimics the observational procedure by decomposing their synthetic
spectra, in which they have incorporated noise and instrumental effects, into
lines that match the spectrum as closely as possible. Under an ideal
observational arrangement, the two approaches should agree. Observers
attempt to ``correct'' for the resulting incompleteness and biases
introduced into the deduced line lists through Monte Carlo simulations
of their analysis procedure (e.g., Hu \etal 1995). We will therefore compare
with the results of the observers including their Monte Carlo corrections
when possible.

The described procedures of line finding and fitting have been fully
automated and combined into a spectral reduction
package which is fast and efficient
enough to analyze thousands of lines in just a few minutes on a workstation.

\subsection{Comparison of Methods}
\label{sec:spec_com}

To test the accuracy of our spectral reduction method, we
compare results from the different analysis techniques
to the same synthetic spectrum.
Figure \ref{fig:spec_com} shows the differences between applying four
variations of the methods to
the same numerical spectrum of length 2500 \kms at $z=3$.  A
transmission
cutoff of $F_t=0.7$ is used in all cases.  The solid lines are the raw spectral
data and the dots represent the identified and fitted lines.
Panel (a) is the threshold method with no deblending and no Voigt profile
fitting.  Panel (b) is the same as (a) except that each of the lines
identified with the threshold method have been fitted to a Voigt profile.
There are a total of 14 lines in this piece of the spectrum without deblending.
Panel (c) shows the identified lines and their fits from our reduction
method with a deblending algorithm, and panel (d) is the same as (c)
except that we use the opacity data instead of the spectral transmission data
to fit the saturated lines in order to obtain more accurate values for the
equivalent width and the column density.  There are a total of 19 lines found
with this deblending scheme.
We see that without deblending a large percentage (approximately $26\%$)
of lines are missed even for this low spectral cutoff.

In Table \ref{tab:spec_com}, we list the properties of several identified
lines found in Figure \ref{fig:spec_com}
and compare and contrast the methods with and without deblending.
Two classes of lines are discussed:  single isolated lines,
and lines that are blended below a transmission cutoff of $F_t=0.7$.
For the unblended lines (\#1, 2, and 3),
the $W_0$ and $\NHI$ differ by about 30--50\% between the two methods,
and it is evident that equations (\ref{eqn:spec_wid}) and (\ref{eqn:spec_int})
give systematically smaller values.
The equivalent width
computed according to equation (\ref{eqn:spec_wid}) is underestimated
to a greater degree for lower spectral cutoffs and for weaker lines
since the threshold method misses the important
contributions to the integral above the cutoff.
For the higher spectral cutoffs, this error would get smaller
but more lines would be missed since there are
many more low column density lines than high column density lines.
For the blended lines (\#4 and 5), the differences in the two analysis
methods are much larger than for the unblended cases.
Line \#4 is actually a blend of three
component lines and line \#5 consists of two component lines.
The individual components are observed to be a single absorption feature
in the threshold method, and the physical parameters of the blended complex
of lines are essentially the sum of the deblended component lines.

\begin{figure}
\plottwo{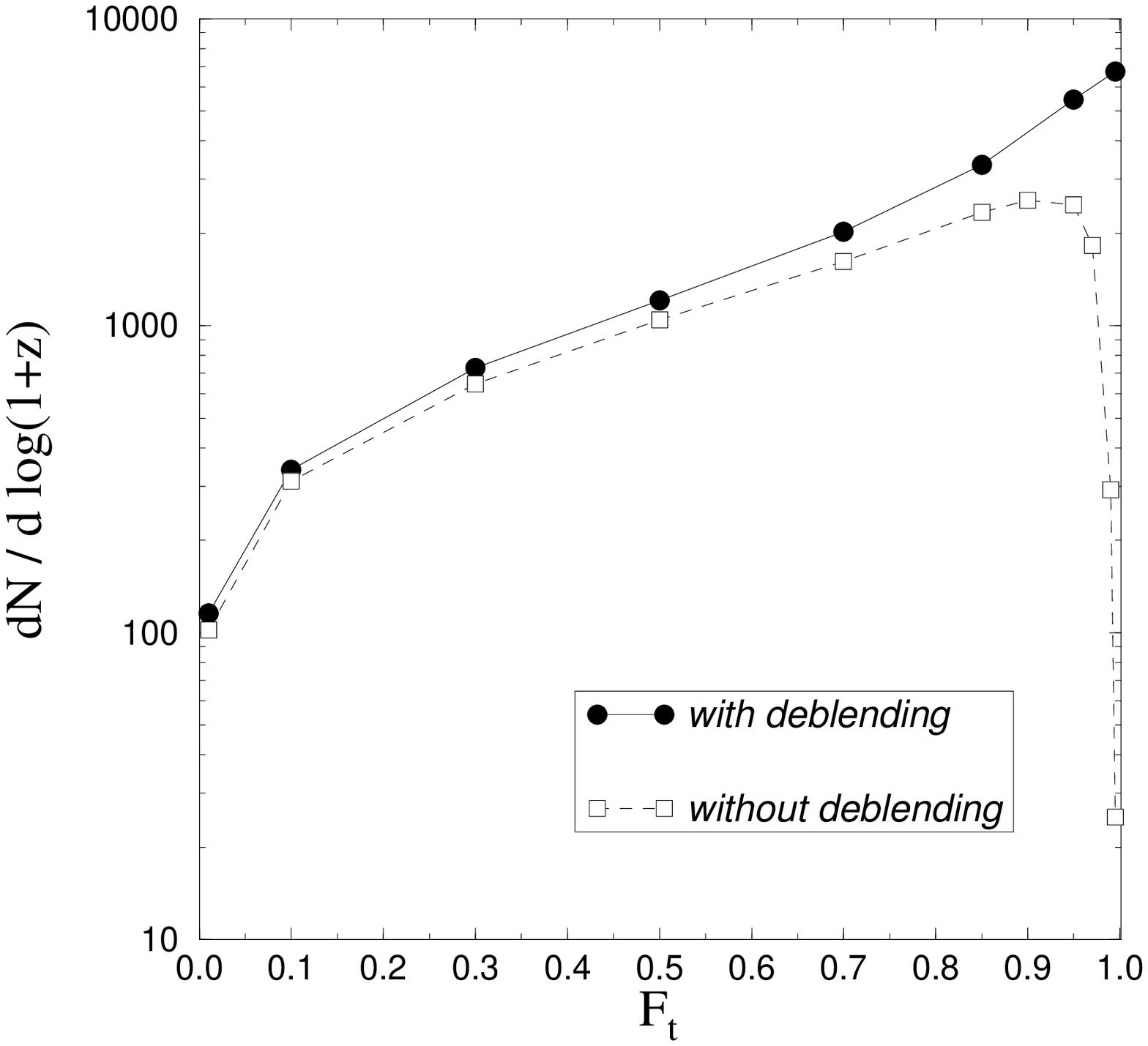}{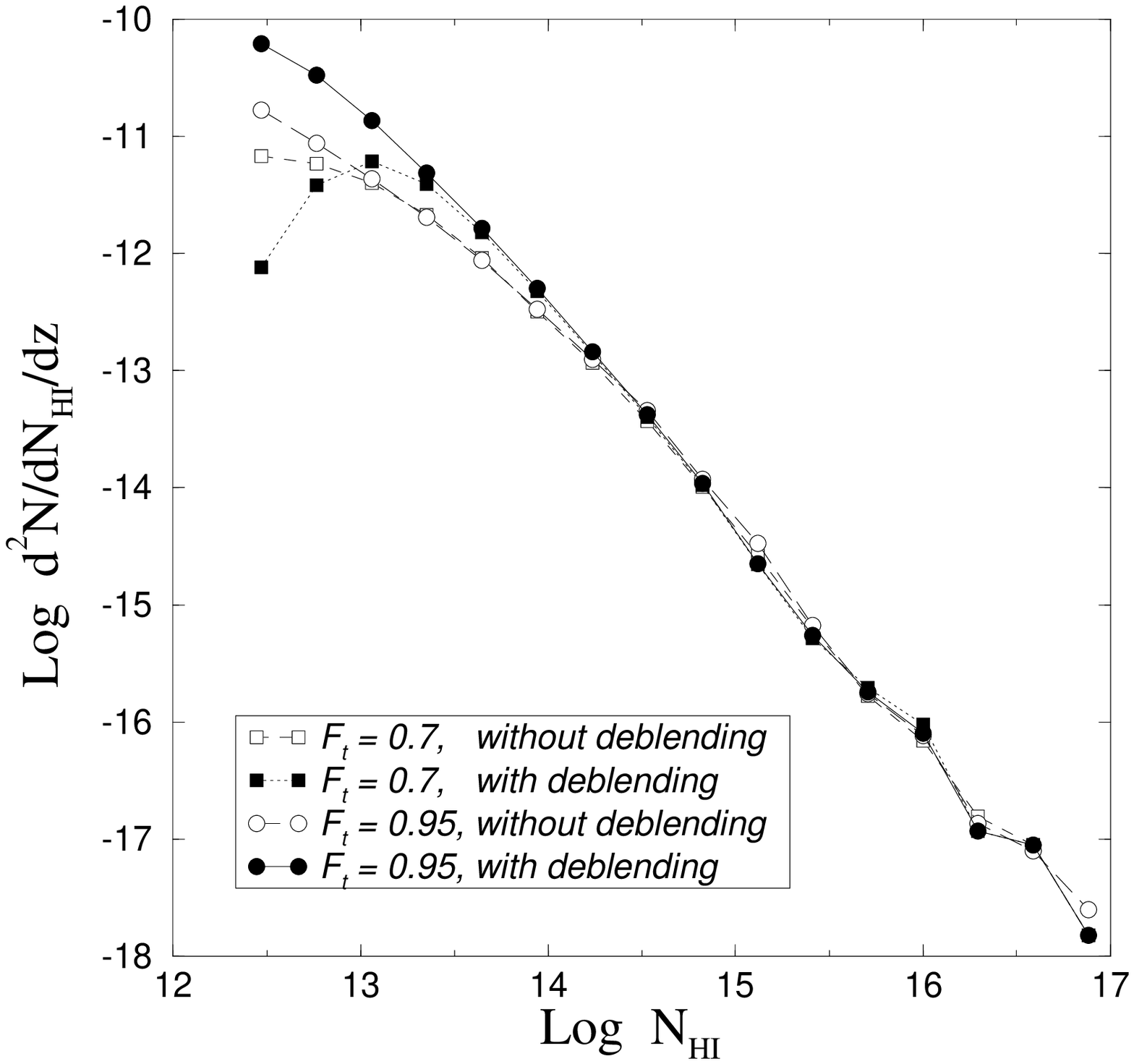}
\caption{ The number of identified lines (per unit $\log (1+z)$ interval)
at $z=3$ in 100 random line--of--sight samples
is plotted as a function of the spectral cutoff $F_t$.
The filled circles are results from the deblending spectral analysis method;
the open squares represent results from the threshold method.
For the typical cutoff $F_t=0.95$, the threshold method misses
about 55\% of the lines.}
\label{fig:spec_cut}
\caption{Column density distributions obtained from the
two spectral analysis methods: with (filled symbols) and without
(open symbols) deblending. Results from 100 random line--of--sight samples
using two different transmission cutoffs,
$F_t=0.7$ (squares) and $F_t=0.95$ (circles), are shown.}
\label{fig:spec_col}
\end{figure}

Figure \ref{fig:spec_cut} shows the logarithm of the
number of lines (per unit redshift) found in 100 random LOS
samples as a function of
spectral cutoff for the two methods; with and without deblending.
We see that even for low transmission
thresholds ($F_t\leq0.7$), approximately $10-30\%$ of lines are missed. For a
threshold of 0.95 corresponding to the signal--to--noise ratio of the Keck
telescope, the total number of identified lines
per unit redshift at $z=3$ is 5455 with deblending and 2471
without deblending, which means that roughly $55\%$ of the lines are missed.
For higher cutoffs, the situation becomes even worse.

To demonstrate the difference between the deblending and threshold
algorithms as well as the effect of different spectral cutoffs, we
plot, in Figure \ref{fig:spec_col}, the column density distributions
for the following four cases: $F_t=0.7$ and $F_t=0.95$, each with and
without deblending.  We see that for $\NHI\geq10^{14}$ \cm2 there are
little differences between the different cutoffs and between the
different reduction methods.  However, at the low column density end,
the differences become large (especially considering that Figure
\ref{fig:spec_col} is a logarithmic plot).  If we take a turn--over in
the power--law dependence of the column density distribution as an
indication of the completeness limit in the distribution, we find that
with deblending and a high cutoff, we are able to achieve completeness
down to $\sim10^{12}$ \cm2. Without deblending, the completeness column
density is more than one order of magnitude higher and deviations from
the deblending scheme are observed as high as $3\times10^{14}$ \cm2.
The lack of low column density lines can significantly affect the
statistical properties of the \Lya forest. For example, the median
Doppler parameter obtained by fitting the Doppler parameters to a
Gaussian distribution is 23.7 \kms~with deblending and 21.8
\kms~without deblending.  Also, the slope of the equivalent width
distribution $W^*$, obtained by fitting to an exponential function, is
$0.11$~\AA~ with deblending and $0.25$~\AA~ without deblending.

Finally, we end this section by noting that observers use several
different spectral reduction methods for the QSO absorption lines.
Some observers rely on their experience to judge whether or not to
deblend a certain line.  Others use some automated package such as
VPFIT (Carswell 1995) which employs an iterative procedure to minimize
the fitting error.  For the best comparison to be made between the
simulations and the observations, it is imperative that the same
procedure be adopted for the analysis of both the simulated and the
observed spectra. For now, we have limited ourselves to extracting the
actual underlying cloud properties that would be measurable under
idealized observational conditions. Certainly the observations are
informing us of the true properties of the clouds over some column density
range, but at some sufficiently low column density the finite resolution and
signal--to--noise of the detectors become limiting factors, as well as
the blending of the absorption lines themselves. We have here assumed
essentially infinite signal--to--noise and resolution for our spectra,
and devised an algorithm for deblending the lines. The deblending is
now essential for comparing with the observations of the forest at
moderate to high redshifts ($z>2$) at the high spectral cutoffs
($F_t>0.9$) that may now be achieved with the high resolution, high
signal--to--noise instruments like the Keck HIRES.

\section{Spectral Properties at Redshift $z=3$}
\label{sec:stat_sta}

In this section, we discuss the statistical properties of the
\Lya absorbers at a fixed redshift $z=3$,
derived from the spectral analysis method described in \S \ref{sec:spec}.
As in observations, a particular spectral cutoff is needed to generate a
line list from the computed opacity.  In order to compare with the
highest available resolution data, we employed the same spectral threshold
as for the Keck HIRES data, $F_t=0.95$.  However, the advantage of numerical
simulations is that an arbitrarily high spectral cutoff can be used to make
predictions for any improved future observations.
It is therefore of interest
to use the highest spectral cutoff possible to identify all the lines
and derive results
from a purely theoretical viewpoint.
Hence, in addition to the HIRES cutoff value, we will also
present results from the line list obtained by using the low
opacity cutoff of $1-F_t = 10^{-5}$,
chosen so that it is lower than the minimum
opacity at the smallest redshift of our data ($z=0.5$).
The line list with this cutoff thus includes all
the generated absorption features.

To create a statistically representative data base, we probe along many
randomly directed lines--of--sight (or samples in our case).
For each redshift interval, $\Delta z = 0.1$, we have generated 913 samples
for \HI and 500 samples for \HeII.  At $z=3$, for example, the
total number of lines in our data set is 73244 for \HI and 159630 for
\HeII. The results presented in the following sections
are based on the statistics from these large databases.

\subsection{Column density distribution}
\label{subsec:colden}
\begin{figure}
\plotone{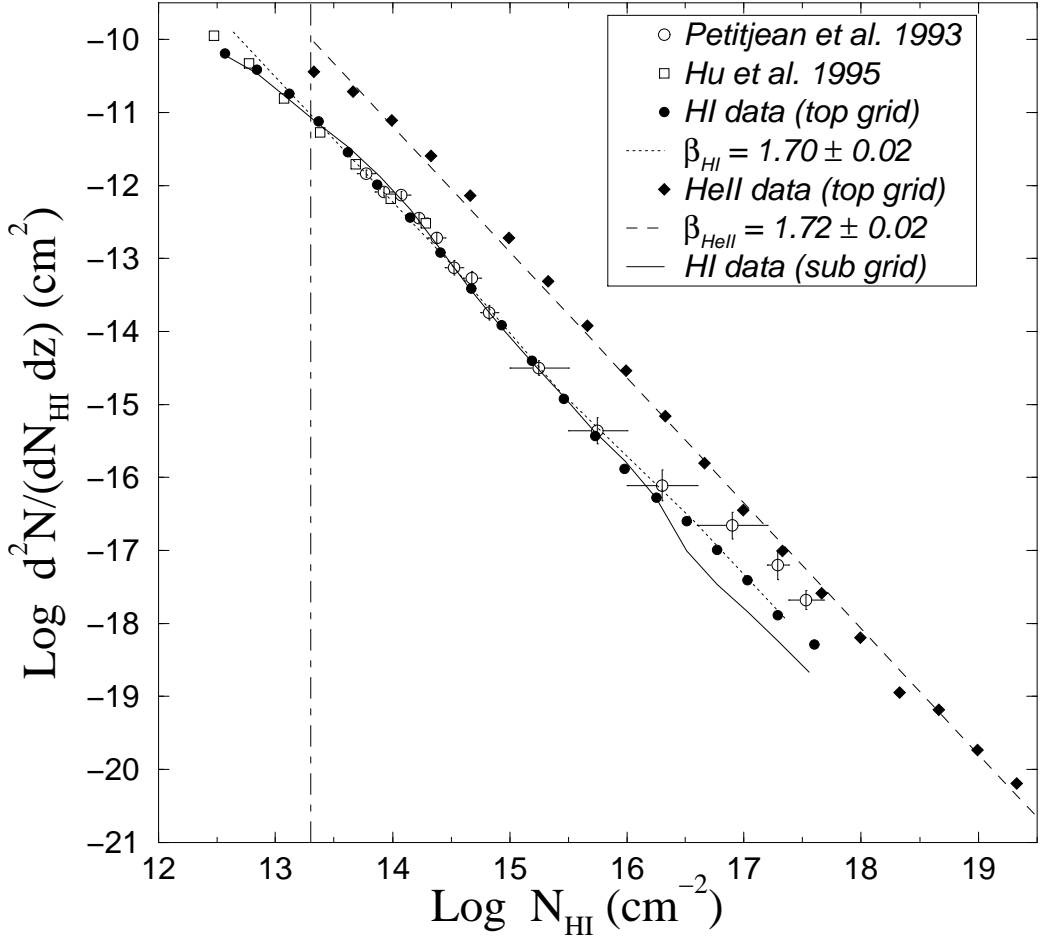}
\caption{\HI (filled circles) and \HeII (filled diamonds)
column density distributions at $z=3$.
Also shown are the observed data from
Petitjean \etal (1993) and Hu \etal (1995).
The dotted and dashed lines are the least--squares fits to the entire
column density ranges of \HI and \HeII respectively.
The \HI data can also be fit at the high and low ends with
$\beta=1.39\pm0.06$ for $2\times10^{12} < N_{\rm HI} < 10^{14}$ \cm2
and $\beta=1.71\pm0.03$ for $10^{14} < N_{\rm HI} < 3\times10^{17}$ \cm2.
The solid line represents the \HI distribution derived from data on
the higher resolution sub grid.
The vertical dashed line indicates
$\NHI=2\times10^{13}$ \cm2, the dividing line between the
optically thin and thick components. For the \HeII distribution,
we find $\beta=1.72\pm0.02$ over the range plotted.}
\label{fig:stat_nlNH_HI}
\end{figure}
The \HI column density distribution is plotted in Figure
\ref{fig:stat_nlNH_HI} along with the Petitjean \etal (1993) and Hu
\etal (1995) (including their correction for incompleteness) observed data.
There is a deficiency of lines
between column densities $10^{15}$ -- $10^{16}$ \cm2, compared to the
extrapolated power--law fit to lower column density systems,
both in the simulation and in the
observed distribution (as first noted by Carswell et al. 1987). This deficiency
is already apparent in the simulation results using the direct method
in ZAN95. The numerical data (filled circles) can be fit to a power
law (dotted line) over the entire range of column densities,
$dN/dN_{\rm HI} \sim \NHI^{-\beta}$ with index $\beta=1.70\pm0.02$.
It is also of interest to fit the distribution separately at the low
and high column density ends. For the \Lya lines with column
densities between $2\times10^{12}$ \cm2 and $10^{14}$ \cm2, the power
law index is found to be $1.39\pm0.06$.  For lines with column
densities between $10^{14}$ and $3\times10^{17}$ \cm2, the index is
$1.71\pm0.03$. These slopes match the Giallongo \etal (1996) results
of 1.4 and 1.8 very well. We are, however, unable to reproduce the observed
data at the highest column densities, $\NHI>~\mbox{few}~\times10^{16}$ \cm2.
This is likely partly due to the
finite computational box size which limits the large scale power, and
the finite resolution of the computation, but also because for column
densities greater than $8 \times 10^{16}$ \cm2, the opacity at the
Lyman edge exceeds 0.5 and self--shielding from the ionizing radiation flux
(which we have not included) becomes important. The corrections due to
self--shielding can be substantial, without which the neutral hydrogen
density can be severely underestimated. Katz et al. (1996) have
included a correction for the opacity; nonetheless, even after making
the correction, these authors similarly find a significant deficit of
high column density systems after matching the normalization to the
low column density end. Hence radiative transfer can account for only
part of the discrepancy, and it is likely that at least part of
the deficit is numerical in origin.

The amplitude of our simulated column density distribution is fixed by
our choice of cosmological parameters and the Haardt \& Madau (1996)
UV radiation field; i.e., no renormalization has been implemented. The
overall agreement with observations is excellent, with deviations on
the order of 50\%. However, the \HI column densities (for fixed $h$),
scale as the ratio of physical parameters $b_{\rm ion}^{\rm HI}
\equiv\Omega_b^2/\Gamma_{\rm HI, -12}$, where $\Gamma_{\rm HI, -12}$ is
the photoionization rate at the \HI Lyman edge in units of
$10^{-12}\, {\rm s^{-1}}$. (The ratio for \HeII may be similarly defined.)
The ratio $b_{\rm ion}$ effectively acts as
an ``ionization bias,'' relating the column density of an absorber to its
total hydrogen (or helium) column density.
This simple scaling arises because neutral hydrogen \HI is created by
the recombination of protons and electrons, which scales as $\rho_b^2
\propto \Omega_b^2$, while also being destroyed by
photoionization from the UV radiation, which scales as $\Gamma_{\rm
HI}$, over periods of time much shorter than the Hubble time. Note
this scaling is valid only to the lowest order when effects such as
collisional ionization can be neglected. It also neglects the density
dependence of the gas temperature. Provided the cosmological
baryon density is not scaled by more than a factor
of $\sim 2$ from the simulation value, the scaling
with $\Omega_b$ is approximately correct. Our choice of model
parameters and radiation intensity results in a fortuitous coincidence
with the measured number density of systems. We find close agreement
with the line density amplitude for $\NHI>10^{14}$\cm2 found by Dav\'e
\etal (1997), after adjusting the two distributions to the same value
of $b_{\rm ion}$. We do, however, find significantly more low column
density systems. We return to this difference below.

Also shown in Figure \ref{fig:stat_nlNH_HI} is the corresponding
\HeII column density distribution (filled diamonds). This
distribution can also be fitted to a power--law, with an index of
$\beta=1.72\pm0.02$ (dashed line). Note again the deficiency of lines
in the \HeII distribution, but in the density range between
$10^{17}$ -- $10^{18}$ \cm2.

Finally, we plot the \HI column density distribution found on
the subgrid of the 9.6 Mpc box (solid line). As mentioned in \S
\ref{sec:simulations}, the more refined subgrid is centered on the
most underdense region in the top grid to resolve the lowest column
density clouds. For densities below $10^{16}$ \cm2, the top and subgrid
results match extremely well, indicating that our results are not
sensitive to the spatial grid resolution down to the completeness
density of $\sim10^{12}$ \cm2. The considerably lower amplitude at the
very high end ($>10^{16}$ \cm2) of the subgrid distribution function
is an artifact of centering the subgrid on a void region, and so lacks the
highest density fluctuations.

\subsection{Doppler parameter distribution}
\label{subsec:doppler}

\begin{figure}
\plotone{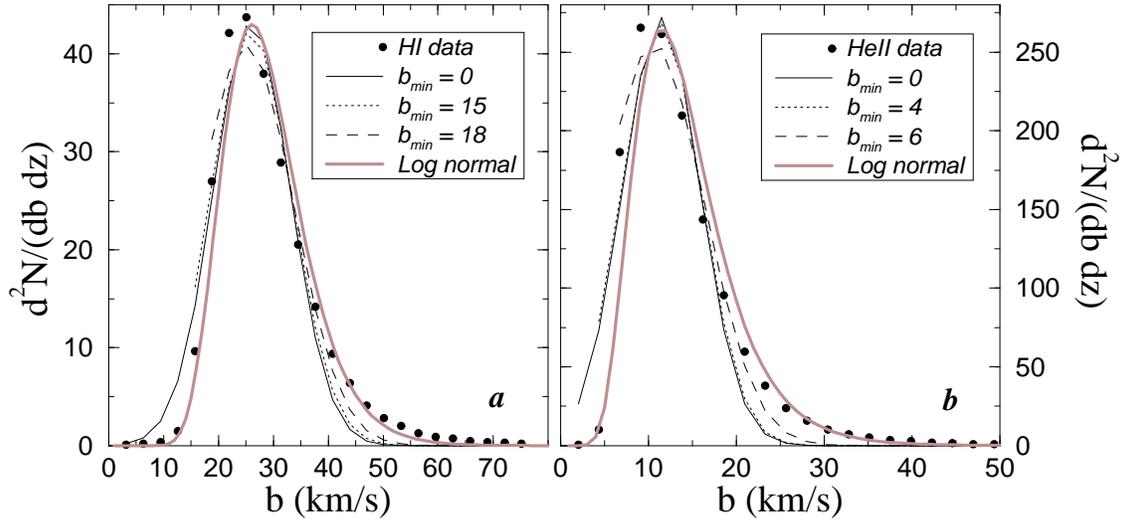}
\caption{Doppler parameter distributions for \HI and \HeII at $z=3$.
The filled circles represent numerical data and the different line types are
various truncated Gaussian fits with the indicated minimum Doppler parameter.
The mean and width for each \HI distribution are
$\bar b=26.1$, 25.9, and 25.1\kms, and $\sigma=7.0$, 7.4, and 8.5\kms
for $b_{\rm min}=0$, 15, and 18\kms, respectively.
These agree closely with the Keck HIRES
values of $\bar b=28$\kms and $\sigma = 10$\kms for $b_{\rm min}=20$\kms
at $z\approx3$ (Hu \etal 1995), and $\bar b=23$\kms and $\sigma=8$\kms for
$b_{\rm min}=15$\kms at $z\approx4$ (Lu \etal 1997). The non--gaussian
tail for large $b$ is well captured by a log normal distribution, with
$\bar b=26.1$ \kms, and dispersion $\sigma_{ln}=0.27$. For \HeII, we find
respectively for $b_{\rm min}=0$, 4, and 6\kms, $\bar b=11.5$, 11.5, and
10.6\kms and $\sigma=4.4$, 4.5, and 5.8\kms. The log normal parameters
are $\bar b=11.5$ \kms and $\sigma_{ln}=0.38$.}
\label{fig:stat_nlb}
\end{figure}

\begin{figure}
\plotone{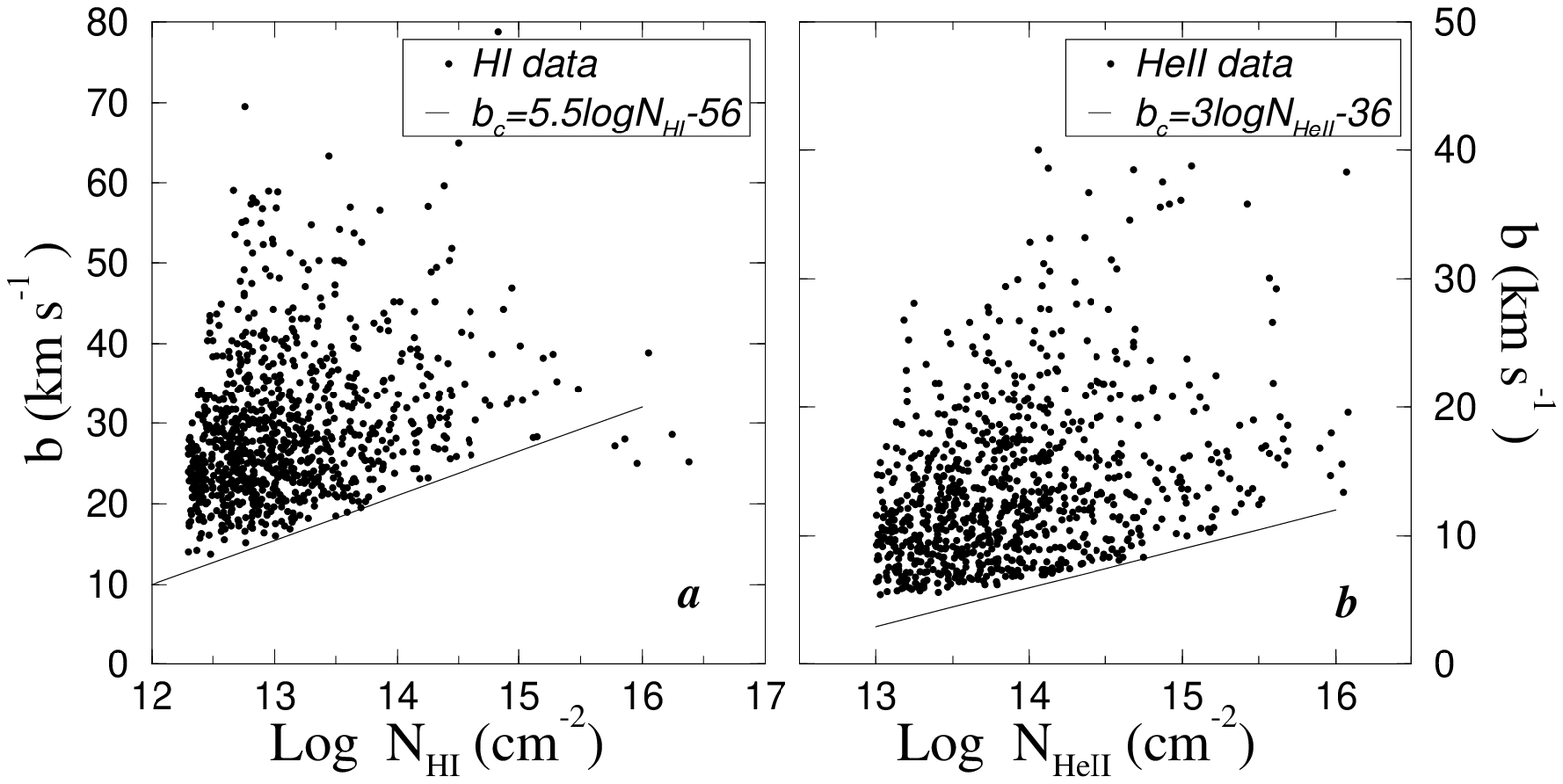}
\caption{Scatter plots of the Doppler parameter versus column density
for both \HI and \HeII at $z=3$. We find no apparent correlation between
$b$ and $N$. The lower cutoff in the $b-\NHI$ plot for
column densities near $10^{14}$ \cm2 is $b_c=15.5 + 5.5 \NHI/10^{14}$,
derived by expanding the expression in the figure legend around
$\NHI \sim 10^{14}$ \cm2. This limit is consistent
with the Hu \etal (1995) result $b_c=4 \NHI/10^{14} + 16$.
The left border is an artifact of setting the threshold for
line detection to $\tau_0 \approx 0.05$, which imposes
$b < 1500 (\NHI/ 10^{14})$ \kms.}
\label{fig:stat_bNH}
\end{figure}

Figure \ref{fig:stat_nlb} shows the \HI and \HeII Doppler parameter
distributions from the simulation. The high resolution Keck HIRES data
reveal a lower cutoff to the Doppler parameter with a possible weak
dependence on column density (Hu et al. 1995; Lu et al. 1997).
Our data exhibit a similar lower cutoff, with a column density dependence
for both the \HI and \HeII data, as shown in Figure
\ref{fig:stat_bNH}. The cutoff for the \HI case is well fit by
$b_c=5.5\log \NHI-56$ and, expanding for column densities
$\NHI \sim 10^{14}$ \cm2, we find $b_c\approx15.5 + 5.5 \NHI/10^{14}$,
which compares favorably to the Hu \etal
result of $b_c\approx16 + 4 \NHI/10^{14}$. The left border is an artifact of
setting the threshold for line detection to correspond to
$\tau_0\approx0.05$, which imposes $b < 1500 (\NHI/ 10^{14})$ \kms
(cf. eq. [\ref{eqn:spec_col}]).

The Doppler parameter obtained from the spectral analysis procedure is
significantly smaller than that found using the direct method
described in ZAN95. There, $b$ was computed according to the physical
state of the clouds by averaging the temperature and the projected LOS
velocity throughout each individual structure. This direct approach is
only a crude approximation. Because the \HI and \HeII absorbing
regions are generally colder than the entire overdense region selected
out through the criteria $\rho/\overline\rho>1$, the Doppler parameter
can be significantly overestimated, especially for the larger
structures. We also note that the Doppler parameter computed from the
more sophisticated deblending analysis procedure yields a lower value
of $b$ than the threshold method. The deblending algorithm is able to
isolate individual lines within otherwise broader blended features,
thereby reducing the line width. A similar difficulty is encountered
in the observations, and the trend has been to find a decreasing mean
Doppler parameter with increasing spectral resolution and
signal--to--noise ratio.

We fit the Doppler parameter distribution to a truncated Gaussian
distribution:\ $f(b) \propto \exp[-(b-\bar b)^2/ 2\sigma^2]$ for
$b>b_{\rm min}$, and $f(b)=0$ for $b<b_{\rm min}$. We show three fits
for \HI in Figure \ref{fig:stat_nlb}, with the Gaussian distribution cut off at
$b_{\rm min} = 0$, 15, and 18 \kms. We obtain for the respective Gaussian
means and widths, $\bar b = 26.1$, 25.9, and 25.1 \kms, and
$\sigma=7.0$, 7.4, and 8.5 \kms. The average Doppler parameter is
nearly independent of the assumed cut-off, while the growth in the
width reflects the improved fit to the distribution. These values
agree well with those of Hu et al. (1995), who, based on Monte Carlo
simulations, find that their data at $z\approx3$ are adequately described by
an underlying truncated Gaussian distribution with $\bar b=28$ \kms~and
$\sigma=10$ \kms~for $b_{\rm min}=20$ \kms. Our results agree somewhat more
closely with those of Lu et al. (1997), who find $\bar b=23$
\kms~and $\sigma=8$ \kms~for $b_{\rm min}=15$ \kms, though these values are
for absorbers at $z\approx4$. We find a non--gaussian tail for
$b>40$ \kms. A tail is present in the Keck HIRES data as well, but may
be due to line-blending, and is reproduced in the Monte Carlo
simulations of Hu et al. and Lu et al. We find a log normal distribution,
$f(b)\propto\exp[-\log^2(b/\bar b)/2\sigma_{ln}^2]$, with $\bar b=26.1$ \kms
and $\sigma_{ln}=0.27$, well matches the tail. It also naturally cuts off
at low $b$ without requiring a truncation in the distribution.
Our mean Doppler parameter and distribution width for the Gaussian
fits are smaller than the values found in
the simulations of Hernquist et al. (1996) (but at $z=2$), for which
the threshold method of line-fitting was adopted. Their results agree
fairly well with the Gaussian fit Press \& Rybicki (1993) inferred from the
equivalent width distribution of Murdoch et al. (1986):\ $\bar b=32$ \kms~and
$\sigma=23$ \kms~for $b_{\rm min}=0$. Even after refitting their lines by Voigt
profiles, Dav\'e et al. find a mean $b$--value and distribution width
at $z=3$ that exceed ours. The FWHM of our distribution is
$\sim20$ \kms, in close agreement with the Keck data (Hu et al. 1995;
Lu et al. 1997). The FWHM of the distribution in Dav\'e et al. is
$\sim30$ \kms (their Figure 3). It is unclear at this point whether the
disagreement is a consequence of the difference in line analysis software
or due to a difference between the simulations.

For the \HeII Doppler parameter distribution, the values we obtain for the fit
parameters with respective cutoffs $b_{\rm min}=0$, 4, and 6 \kms,
are $\bar b=11.5$, 11.5, and 10.6 \kms, and $\sigma=4.4$, 4.5, and 5.8 \kms.
The mean Doppler parameters are slightly smaller than would be given directly
by the scaling of the thermal values, $(m_{\rm He} / m_{\rm H})^{-1/2}=0.5$.
The reason is that we have used the same threshold cut at $F_t=0.95$ for
the \HeII spectra as used for \HI. The distribution thus includes, and in
fact is dominated by, systems with much lower \HI column densities than was
used in the \HI $b$--distribution fit above. These low column density
systems appear to be somewhat cooler than their higher column density
counterparts. This trend is apparent in Figure \ref{fig:stat_bNH}, which
also shows that $b_c$ for \HeII is slightly smaller than given by the
thermal scaling from the value for $b_c$ found for the \HI at the higher \HI
column densities. In the region of overlap $\log N_{\rm HI}>12.5$ and
$\log N_{\rm HeII}>14.1$ (using $\NHeII/\NHI\approx40$ at $z=3$ for the
Haardt \& Madau spectrum), the ratio of $b_c$'s is consistent with
thermal scaling. The corresponding temperature is $T\simeq1.3\times10^4\kel$.
As for the \HI, a non--gaussian tail is present. It is well fit by
a log normal distribution, with $\bar b=11.5$ \kms and $\sigma_{ln}=0.38$.

Pettini (1990) had suggested a possible correlation between the Doppler
parameter and column density, but this has not been confirmed by
other observers. Figure \ref{fig:stat_bNH} shows the $b-N$
scatter plots for \HI and \HeII.  Other than for the weak column density
dependent lower cutoff in the distribution, there is no apparent correlation
in the data. This result is also consistent with the conclusions
drawn from the direct method analysis in ZAN95.
As shown in ZAN95, the scatter in the data
is due, in part, to the broadening of lines from the bulk velocity of
the baryonic fluid. This is evident in the low density end, where the gas is
at about the same uniform temperature $\sim 1$ \ev. The dispersion of
points at low column densities in Figure \ref{fig:stat_bNH} thus reflects
the range of velocities associated with the absorber fluctuations.

\subsection{Equivalent width distribution}
\label{subsec:eqw}

Observations show that the rest-frame equivalent width distribution can be fit
to an exponential relation: $dN/dW_0=(N^*/ W^*)e^{-W_0/W^*}$, with $W^*$
in the range $0.25 - 0.35$~\AA~ for the medium resolution observations
(Sargent \etal 1980; Murdoch \etal 1986), and in the range
$0.09 - 0.15$~\AA~ for higher resolution observations (Kulkarni \etal 1996).
Overall, our numerical results tend to agree with the higher resolution data,
which show a steeper equivalent width distribution.
\begin{figure}
\plottwo{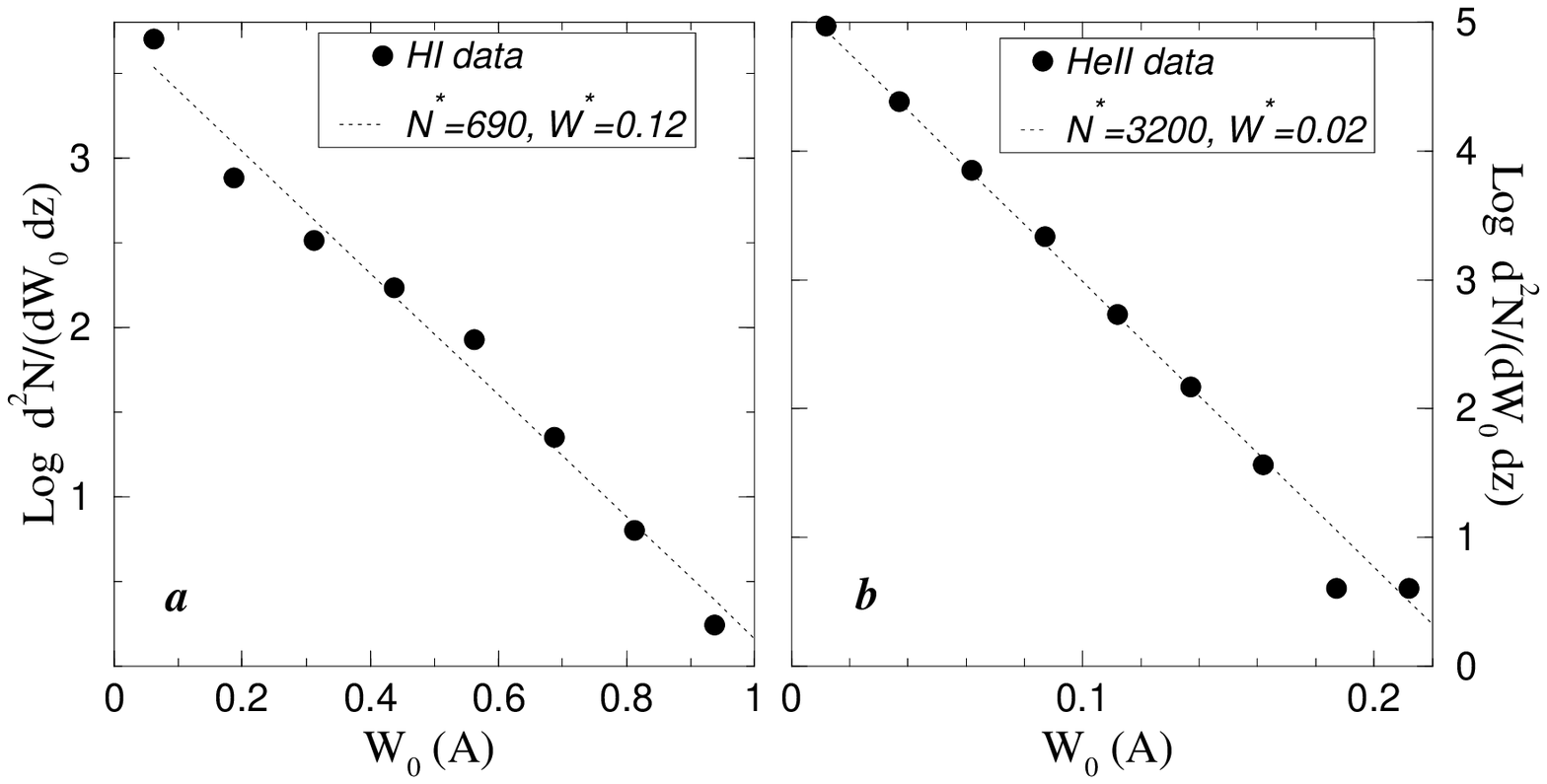}{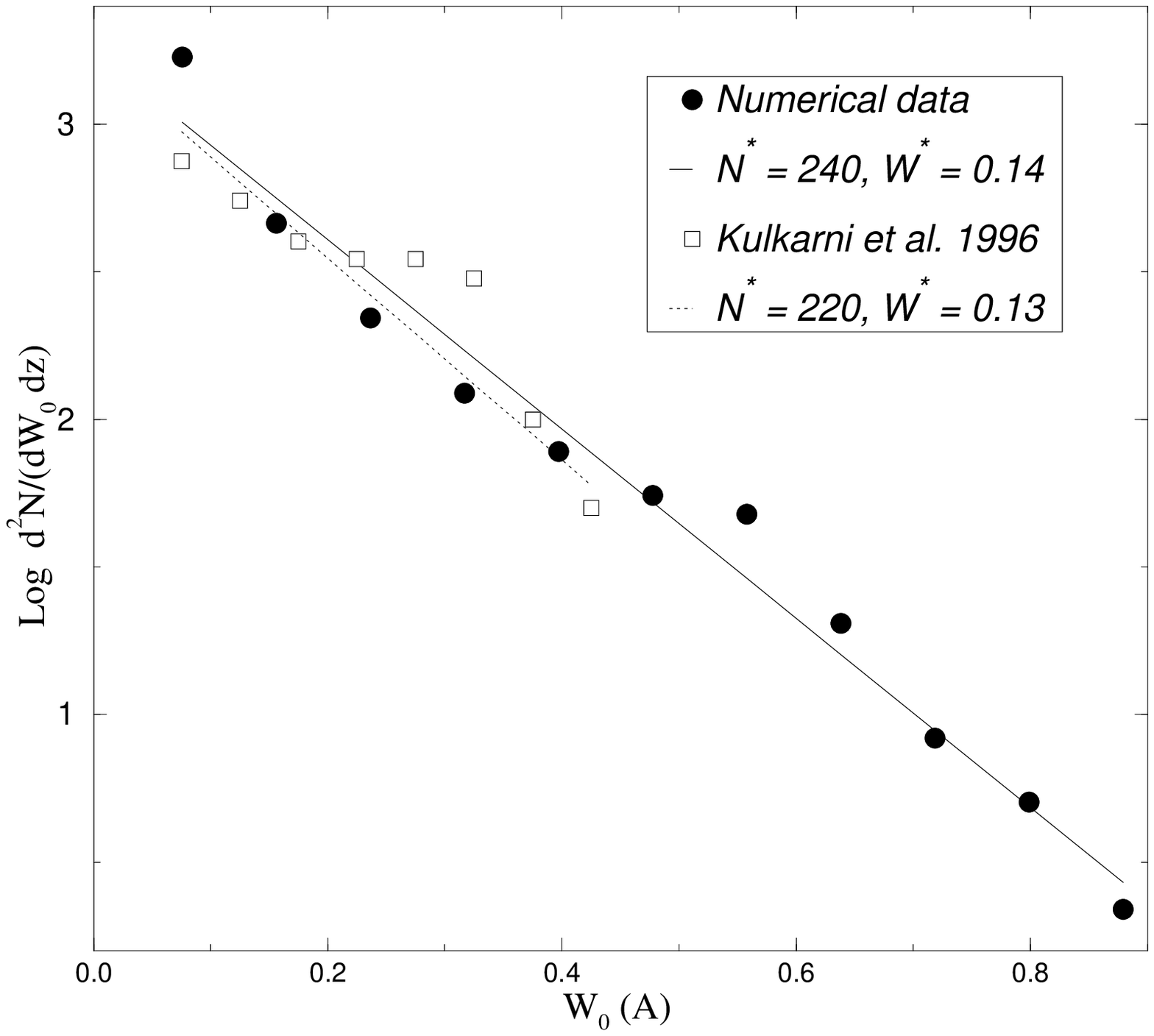}
\caption{Equivalent width distributions for \HI and \HeII at
redshift $z=3$. The filled circles represent the numerical data, and the
dotted lines are fits to an exponential distribution of the form
$dN/dW_0 = (N^*/W^*)e^{-W_0/W^*}$.
The \HI results compare favorably with the Keck HIRES data:
$W^* = 0.17\pm0.02$\,A\, and $N^* = 424$, determined by fitting
an exponential to the data of Hu et al. (1995).}
\label{fig:stat_nlew}
\caption{\HI equivalent width distribution at $z=2$.
The solid line is the exponential fit to our numerical data, which are
represented by filled circles. The dotted line  is the exponential
fit for the KPNO Q1331+170 data (Kulkarni \etal 1996),
indicated by the open squares.}
\label{fig:stat_nlew_z2}
\end{figure}

Figure \ref{fig:stat_nlew} shows the equivalent width distributions
and their exponential fits for both the \HI and \HeII numerical
data. We fit an exponential to the data of Hu \etal (1995)
for $W_0>0.1$\,\AA, and requiring $\NHI>10^{13.2}$\cm2 (their completeness
limit), obtaining $W^*=0.17\pm0.02$\,\AA\, (95\% confidence), and $N^*=424$
(for an average redshift of $z=2.87$). The value $W^*=0.12$~\AA~ from the
simulation is somewhat smaller, possibly due to missing the higher column
density systems as discussed in \S \ref{subsec:colden} above, but it
may be due to incompleteness at low column densities in the HIRES
data as well. The line densities of the two distributions agree well.
The distribution for \HeII is much steeper, with an exponent $W^*=0.02$~\AA~
and amplitude $N^*=3200$, which is much higher than the \HI case.
The \HeII lines are about 4 times more numerous than the \HI lines
above their characteristic equivalent widths $W^*$.

We also plot the equivalent width distribution at redshift $z=2$
along with the KPNO observation
of QSO Q1331+170 at $z\sim1.9$ (Kulkarni \etal 1996)
in Figure \ref{fig:stat_nlew_z2}.
The exponential fits to the two data sets, and the corresponding
parameters are also displayed.
The agreement is excellent in both slope $W^*$ and amplitude $N^*$.

\subsection{Line center opacity}

\begin{figure}
\plottwo{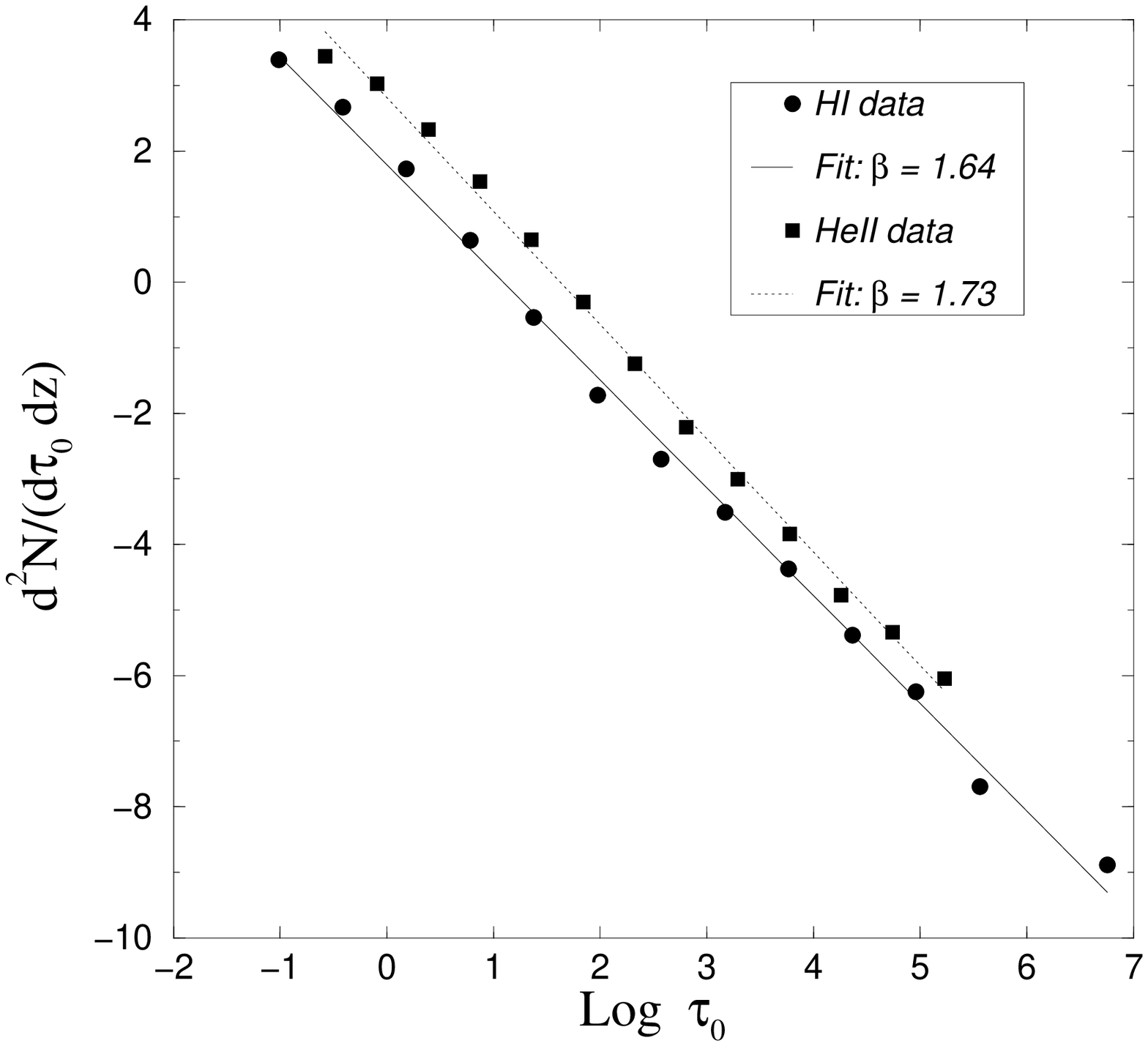}{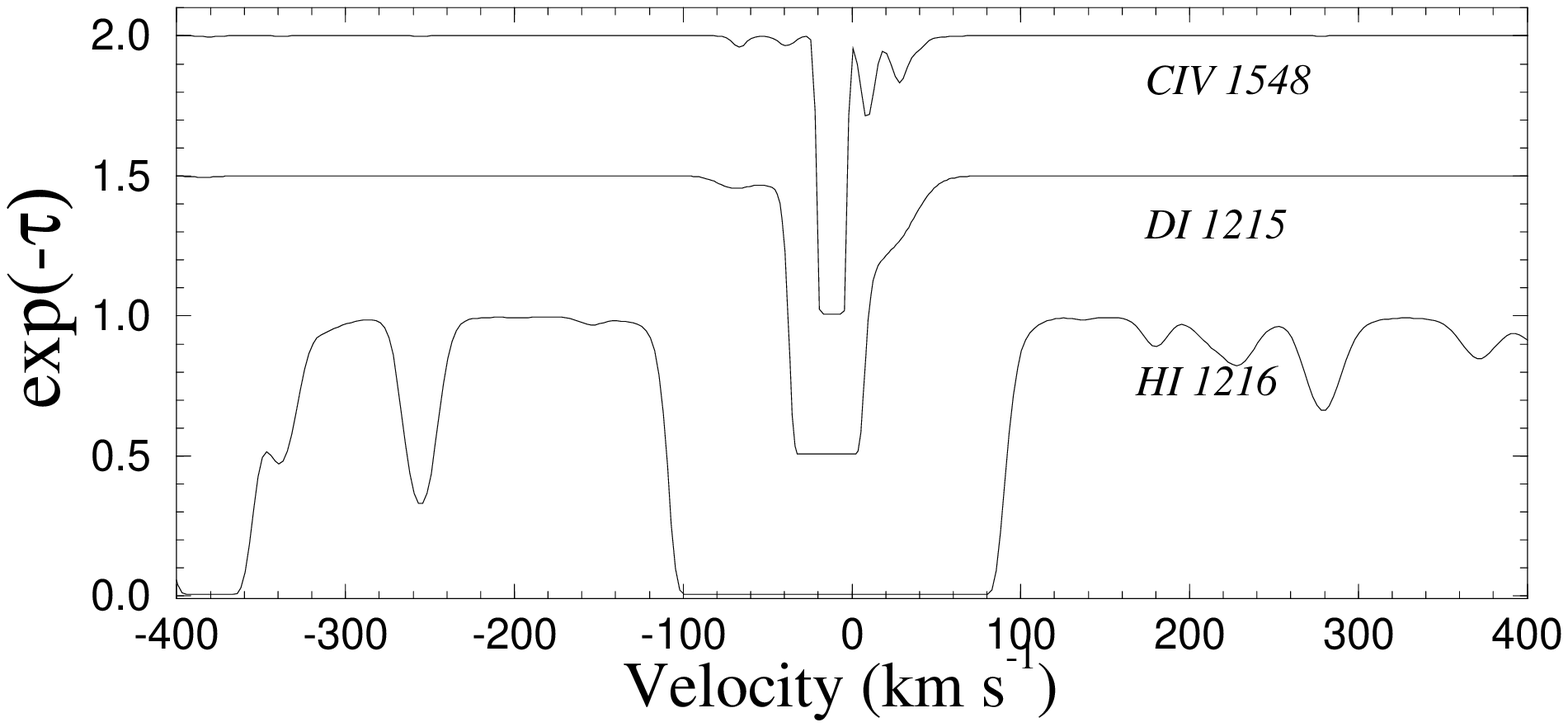}
\caption{Distributions of the line center opacity for \HI and \HeII
at $z=3$, and their corresponding power--law fits.
The overall shape, deficiency of lines, and fitting exponents
are similar to the column density distribution of Figure
\protect\ref{fig:stat_nlNH_HI}.}
\label{fig:stat_nltlc}
\caption{Characteristic spectra of the \DI and \CIV absorption line
systems. The fractional abundances (relative to neutral hydrogen)
are assumed to be $5\times 10^{-4}$ (set artificially high for illustration
purposes) and $2\times 10^{-3}$ for \DI and \CIV respectively, in accordance
with the measurements of Cowie \etal (1995).
To prevent the lines from overlapping, the flux transmissions
are offset by 0.5 and 1.0 for \DI and \CIV, respectively.}
\label{fig:stat_metal}
\end{figure}

We also find the line center opacity can be fit to a power--law
distribution of the form $f(\tau_0)\propto{\tau_0}^{-\beta}$, with
exponents similar to the column density distribution: $\beta = 1.64$
and 1.73 for \HI and \HeII respectively. Figure \ref{fig:stat_nltlc}
shows the \HI and \HeII line center opacity distributions and their
respective power-law fits.

Note there is also a slight deficiency of
lines with respect to the power-law distribution at
$\log\tau_0\sim2$ for \HI and $\log\tau_0\sim3$ for \HeII, which
is consistent with the curvature in the column density distribution.
The similarity between the column density and line center opacity
distributions is understood by noting that $\tau_0$ is proportional to
column density and inversely proportional to the Doppler parameter
(cf. eq. [\ref{eqn:spec_col}]).  Since the line center opacities
and column densities both span about 6 to 8 orders of magnitude, while
the Doppler parameter only changes by factors of less than 10, the
line center opacity $\tau_0$ is roughly proportional to $N_{\rm HI}$
with the transition from $\tau_0<1$ (optically thin) to $\tau_0>1$
(optically thick) occurring at a column density of $\log
\NHI\approx13.3$.

\subsection{Metal line systems}

An uncertainty that has long-existed in the interpretation of the \Lya
forest clouds has been their discreteness. Does a single feature correspond
to a single cloud, or are there subcomponents obscured by the finite
width of the lines? It has recently become possible to investigate
this question observationally by searching for metal absorption in the
forest. Since the width of an absorption feature decreases for higher
atomic masses when the broadening is predominantly thermal in nature,
metal absorption features may act as probes of the internal
structure of the forest systems. We explore this possibility using our
simulations. Since we do not include any elements other than hydrogen
and helium in our simulations, we rescale the hydrogen data by
assuming a uniform fractional abundance of the particular element used
to generate the spectrum. While the ionization fractions will generally
vary with the internal total gas density of the systems even for a
uniform metallicity (e.g., Rauch \etal 1996), here we use the heavier
elements simply to indicate substructure
within a given hydrogen absorption feature.
We consider two heavy elements: deuterium
(\DI $\lambda 1215$) and carbon (\CIV $\lambda 1548$) with abundances
$n_{\rm DI}/n_{\rm HI} = 5\times 10^{-4}$ (set artificially high for
illustration purposes), and $n_{\rm CIV}/n_{\rm HI} = 2\times10^{-3}$,
in accordance with the measurements of Cowie \etal (1995). Figure
\ref{fig:stat_metal} shows an example of the \DI and \CIV lines
located within a saturated \HI \Lya line. The effectiveness of large
atomic mass elements to reveal and probe substructure within the
broader \HI feature is clear. In particular, we point out that the
carbon absorption lines indicate a triplet structure which is a common
feature in observed metal line systems (Cowie \etal 1995).

\section{Evolution Properties}
\label{sec:stat_evo}

\subsection{Line Number}
\label{subsec:linenumber}

Observations have shown that the number of lines
larger than a given column density increases with redshift
as a power law $dN/dz \propto (1+z)^{\gamma}$.
The exact value of $\gamma$ in numerical simulations will reflect
changes in the ionizing radiation, threshold densities
or opacities, merging of structures, the expansion rate of the universe,
intrinsic line deblending with redshift, and the dependence of the
comoving length scale with redshift interval.
We may express the evolution in number density of the clouds intercepted
along a line-of-sight, above a fixed column density threshold, as
\begin{equation}
\frac{dN}{dz}=\frac{c}{H_0\lambda_{\rm eff, 0}}(1+z)^{1/2}\left(\frac{\Gamma}
{\Gamma_0}\right)^{-(\beta-1)}(1+z)^s, \label{eq:dNdz}
\end{equation}
where $\lambda_{\rm eff, 0}$ is an effective comoving mean free path
between absorbers (scaling like the inverse product of the comoving number
density and geometric cross-section of the clouds). The factor
$(1+z)^{1/2}$ includes the $(1+z)^3$ dilution of the proper number
density of the clouds as the universe expands, and the change in the
proper length element with redshift (eq. [\ref{eqn:spec_dz}] with
$\Omega_0=1$ and $\Omega_\Lambda=0$). The term $(1+z)^s$ accounts for
the evolution in the comoving number density and cross-section of
systems corresponding to the column density threshold. For a fixed
comoving number density and cross-section, $s=-2$. The
$\Gamma$--dependent factor accounts for the changing column density
due to evolution of the radiation field, normalized to the
photoionization rate $\Gamma_0$ at $z=0$. A power-law column density
distribution is assumed with exponent $\beta$.

\begin{figure}
\plotone{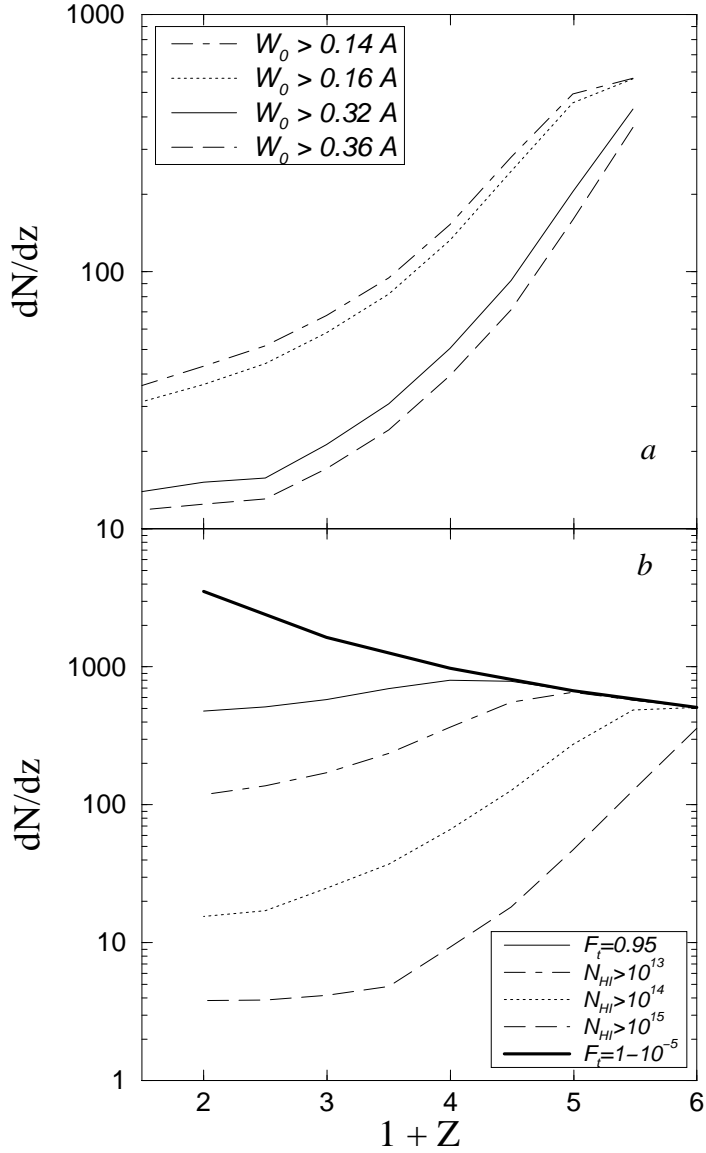}
\caption{Evolution of the number of \HI lines with equivalent widths (a)
and column densities (b) larger than the displayed values.
The lines are identified using a
transmission cutoff $F_t = 0.95$. In the column density plot,
the total line number with this threshold is shown by the solid line,
and the thick solid line is the total number of lines obtained with the
minimum opacity cutoff $1-F_t=10^{-5}$, and thus counts all the generated
absorption lines, including those that fall above $F_t=0.95$.
The thick solid line can be fit to a power--law
$\propto (1+z)^\gamma$ with $\gamma \sim -1.77$.}
\label{fig:stat_nlz_NH}
\end{figure}

Figure \ref{fig:stat_nlz_NH}b shows the evolution of the number
of lines (per unit redshift) with different column density cutoffs.
Also shown are the total line count for the transmission cutoff
$F_t = 0.95$ (solid line), and the total line count for the minimum
opacity cutoff $1-F_t = 10^{-5}$ (thick solid line) which represents all
the generated lines in the simulation. The thick solid line follows a
power--law evolution of the form $\propto (1+z)^\gamma$ with
$\gamma\approx -1.77\pm0.04$, indicating that the complete
absorption line list continues to increase with decreasing redshift.
For the complete line list, the $\Gamma$--dependence is suppressed (all
the lines have been counted), and eq. (\ref{eq:dNdz}) shows that the
clouds evolve with a nearly constant mean free path in the comoving frame, for
which $\gamma=-1.5$. The additional factor may be due to the intrinsic blending
of spectral features at higher redshifts, which effectively increases
the number of lines towards lower redshifts.

There are three notable features in the $F_t = 0.95$ evolution curves:\ the
duration with which each limiting curve follows the thick solid line
increases with smaller density thresholds;
the line number count evolves more rapidly at the higher redshifts;
and the higher column density clouds exhibit a higher rate of decrease.
All three behaviors can be attributed to the evolving radiation field
(cf. Figure \ref{fig:HMradiation}), the fixed transmission cutoff,
and the expansion of the universe.
For example, at $z \gsim 5$ when the radiation field is low,
there is little difference in the number of lines indicated by the
different column density thresholds. Thus most of the clouds
have column densities above $10^{15}$ \cm2. From $z\sim 5$ to $\sim 2$,
the exponential decrease in cloud number reflects the exponential
rise in the radiation field. The departure redshift from the
thick solid line in Figure \ref{fig:stat_nlz_NH} is determined by the
redshift at which the characteristic line center opacity drops below
the threshold opacity associated with the column density cutoffs.
Hence, for a fixed transmission cutoff, the exponential cloud number
dissipation is triggered at smaller redshifts for the lower density clouds,
and since the radiation field evolves less rapidly at the smaller redshifts,
the rate of cloud dissipation is less for the lower density thresholds.
The expansion of the universe also contributes to the dissipation rate,
since the expansion reduces the mean column density and opacity of the IGM.
For a fixed transmission cutoff, the number of lines which fall below the
threshold increases as the universe expands.

Figure \ref{fig:stat_nlz_NH}a also plots the evolution
of the number of lines (per unit redshift) for the different rest frame
equivalent width ($W_0$) cutoffs used by observers
(cf. Table \ref{tab:stat_ew}). The characteristic trends found
in the column density evolutions are also evident here. Namely,
the line number evolves faster in the higher $W_0$ cutoff cases, and
$dN/dz$ for a fixed $W_0$ cutoff evolves faster at
the higher redshifts. Both effects can be attributed, for the most part,
to the changing radiation field.

Table \ref{tab:stat_ew} lists the evolution exponents
$\gamma$ calculated for various $W_0$ limits
and redshift ranges, comparing them with the observed values.
Note the excellent agreement in each case, except for the high
redshift data of Bechtold (1994).
To demonstrate the dependence of the evolution exponent on redshift,
we compute $\gamma$ at three separate redshift intervals for clouds
with equivalent width cutoff $W_0 > 0.32$~\AA. We find
$\gamma=7.70\pm0.09$ for $3.5<z<4.5$,
$\gamma=2.95\pm0.40$ for $1.5<z<3.5$, and
$\gamma=0.24\pm0.03$ for $0.5<z<1.5$.
The evolution in our data is much stronger than that indicated by the
Bechtold (1994) data, although is only somewhat more rapid than the evolution
found using the Keck HIRES data, which we've determined
by performing a maximum likelihood fit to the combined
line lists of Hu et al. (1995) and Lu et al. (1997). The
evolution, however, is sensitive to the evolution of the ionizing UV radiation
field, which becomes increasingly uncertain for $z>3.5$. The evolution would
be slowed if the UV radiation field declines somewhat less rapidly than given
by the Haardt \& Madau spectrum, as would be the case if high redshift QSOs
were obscured by dust in intervening systems (Fall \& Pei 1996). We note that
the enhanced rate of evolution we find at higher redshifts agrees with
the analysis of Zuo \& Lu (1993) for the intrinsic rate of evolution of the
forest, who presume the average \Lya absorption properties arise entirely due
to line--blanketing. They suggest a broken power--law for the increase in
line density, with a break at $z=3.11$, and $\gamma_{\rm int}=2.82$ below
the break and $\gamma_{\rm int}=5.07$ above.

\begin{figure}
\plotone{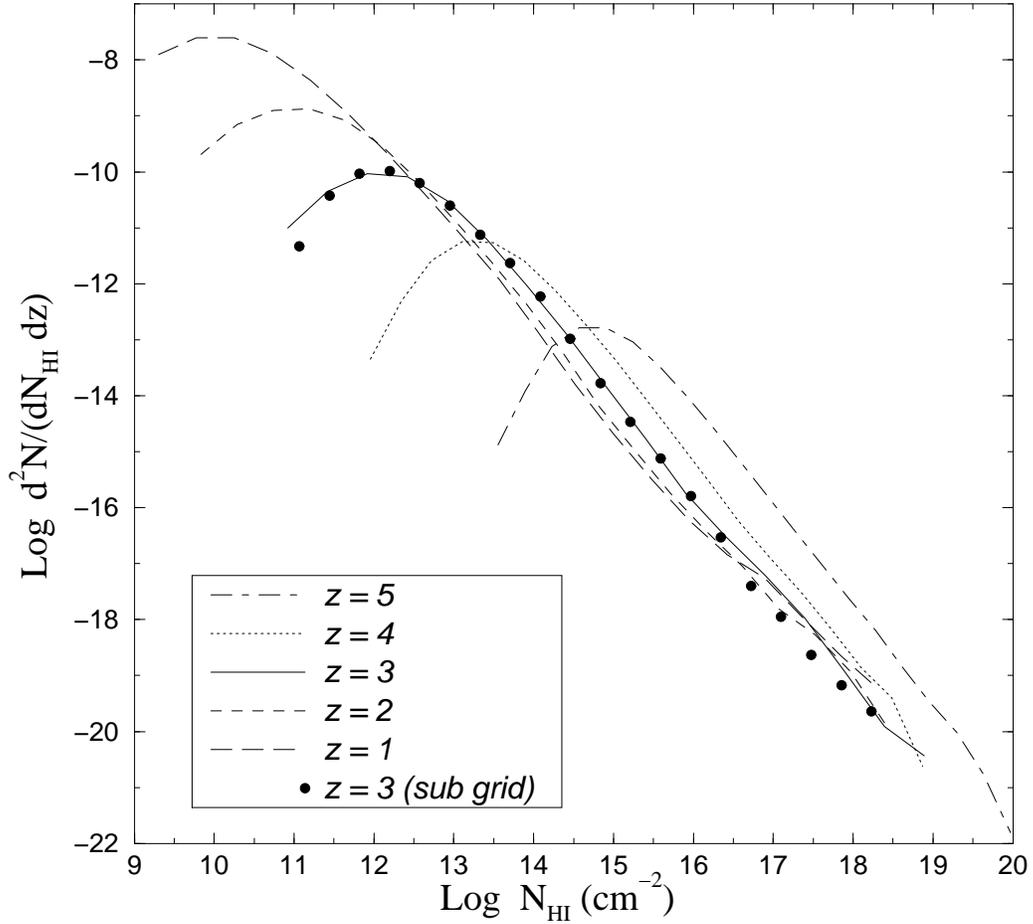}
\caption{Column density distribution at five different redshifts.
The minimum opacity cutoff $1-F_t=10^{-5}$ is used
to identify all the generated lines in the spectra. The turnover
in each curve is an indication of the incompleteness column density.
For our grid resolutions, intrinsic line blending
is the dominant cause of incompleteness
at high redshifts as verified by the sub grid distribution at $z=3$.}
\label{fig:turnover}
\end{figure}

A sequence of column density distributions is shown in
Figure \ref{fig:turnover} at five different redshifts.
For this figure, the minimum opacity cutoff of $10^{-5}$ was used to identify
all the absorption features. Because we do not include radiative transfer,
the statistics of the distribution for $N_{\rm HI}>10^{17}$ \cm2 are not
directly comparable to observations. We do note, however, that the highest
column density systems tend to be associated with the more compact structures.
In particular, the damped systems with $N_{\rm HI}>10^{20}$ \cm2 at $z=5$ are
associated with dense knots in the gas distribution.

In addition to showing the dissipation and
evolution of the clouds, Figure \ref{fig:turnover}
also demonstrates the resolution limitations of our simulations.
The turnovers in the distributions indicate the lower density ends
at which our numerical absorption line lists become incomplete.
This is ultimately tied to the degree of line blending in the
spectra (which becomes more severe at higher redshifts) and the cell size of
the computational grid. For example, we can estimate the minimum
column density resolved in our simulations as the integrated
density in a single cell found in the void regions
\begin{equation}
N_{\rm HI,min} = \Omega_b~f_{\rm HI}~\Delta x_{c}~
\frac{\rho_b}{\overline\rho_b}~\frac{\rho_c}{1.4 m_{\rm p}}~(1+z)^2 ,
\label{grid_resolution}
\end{equation}
where $\Omega_b$ is the baryon fraction, $f_{\rm HI}$ the neutral hydrogen
fraction, $\Delta x_{c}$ the comoving cell size,
$\rho_b/\overline\rho_b$ the baryonic overdensity in the voids,
and $\rho_c/1.4 m_{\rm p} = (3/8\pi)H_0^2/(1.4Gm_{\rm p})$ is the critical
comoving number density of hydrogen atoms in a flat expanding universe.
For our model parameters, a characteristic underdensity of 0.1 for the voids
in which the lowest column density clouds are found (Zhang \etal 1997), and
a typical neutral fraction of a ${\rm few\,}\times10^{-6}$,
we find $N_{\rm HI,min} \sim 10 ^{11}$ \cm2
at redshift $z=3$, somewhat smaller than the completeness limit
of $\sim 10^{12.5}$ \cm2.
We also plot in Figure \ref{fig:turnover}
the column density distribution found in the subgrid evolution
at the redshift $z=3$, using the same minimum opacity cutoff as the top grid.
The distributions match nicely and become
incomplete at the same low column densities,
suggesting that line blending is the dominant source of incompleteness
(for our grid resolutions). A second indication that the turnover is due
to incompleteness in the line-identification comes from the \HeII distribution.
The \HeII distribution turns over only at $\NHeII < 10^{13.3}$ \cm2. For the
Haardt \& Madau radiation field, $\NHeII/\NHI\approx40$ at $z=3$, so that the
\HI distribution should not turnover until $\NHI<10^{11.7}$ \cm2, nearly an
order of magnitude smaller than the turnover in Figure \ref{fig:turnover}.
It is nonetheless possible that the turnovers are partly a reflection of the
minimum column density resolvable by the simulation as estimated by
equation (\ref{grid_resolution}). This may
particularly be true for \HeII, suggesting that the determination of the
underlying $\NHeII$ distribution is at the computational limit.

The distributions in Figure \ref{fig:turnover} extend somewhat below
those of Dav\'e et al. (1997). The difference is particularly
noticeable at $z=2$, where we find the column density distribution
continues to climb to $10^{11}$ \cm2, more than an order of magnitude
smaller than the column density at the peak in their distribution, after
accounting for the differences in baryon density and radiation field between
the two simulations. We cannot offer a definite explanation for the
discrepancy at this point, but it may be due to the differences in the
synthesis of the spectra and in the
line--finding and fitting algorithms. In particular, since Dav\'e et al.
normalize to a higher \HI opacity than we do (corresponding to nearly
a factor of 2 increase in \HI column density for the same physical system
compared to our normalization), the difference is likely due in part
to blending at low column densities. The analysis approach of Dav\'e et al.
is also designed to mimic observational systematics, like noise and the
difficulty in estimating a local continuum level, which would lead them to
reject the weakest lines. In this sense, the difference may indicate the effect
of observational systematics on determing the underlying line distribution.
It may, however, also reflect a
difference in the effective resolutions of the two codes, particularly
in underdense regions where we find most of the low column density
systems reside, or of the numerical algorithms themselves. We return
to this point in Zhang et al. (1997).

\subsection{Effective opacity}
\label{subsec:efftau}

In spectra of insufficient resolution and signal--to--noise
to resolve individual absorption features, the absorption due to the
forest will result from the stochastic overlapping of individual lines.
If the column density distribution extends down to arbitrarily small
values, then the study of absorption by the forest in any spectrum will
be resolution--limited at some level. A comparison between the amount of
absorption expected from the \Lya forest and the actual amount measured
can be used to place a limit on the amount of absorption due to an
underlying undetected component, and has been used to set limits on the
Gunn-Peterson effect (Steidel \& Sargent 1987; Jenkins \& Ostriker 1991).
It is also an essential quantity for assessing the amount of absorption
expected from \HeII given the observed properties of the \HI
\Lya forest (Jakobsen \etal 1994; Madau \& Meiksin 1994). It is thus of
interest to compute the effective opacity in our simulation. The effective
opacity due to line blanketing by clouds Poisson distributed in redshift
can be calculated from the absorption line distribution $d^2N/dW_0dz$ as
\begin{equation}
  \tau_{\rm eff} = \frac{1+z}{\lambda_0}\int{W_0\frac{d^2N}{dW_0dz}dW_0}~,
\label{eqn:mod_teff}
\end{equation}
at redshift $z$ and \Lya wavelength at rest $\lambda_0$
(e.g., Press \etal 1993). The flux depression averaged over all
lines--of--sight
is then $e^{-\tau_{\rm eff}}$.

\begin{figure}
\plottwo{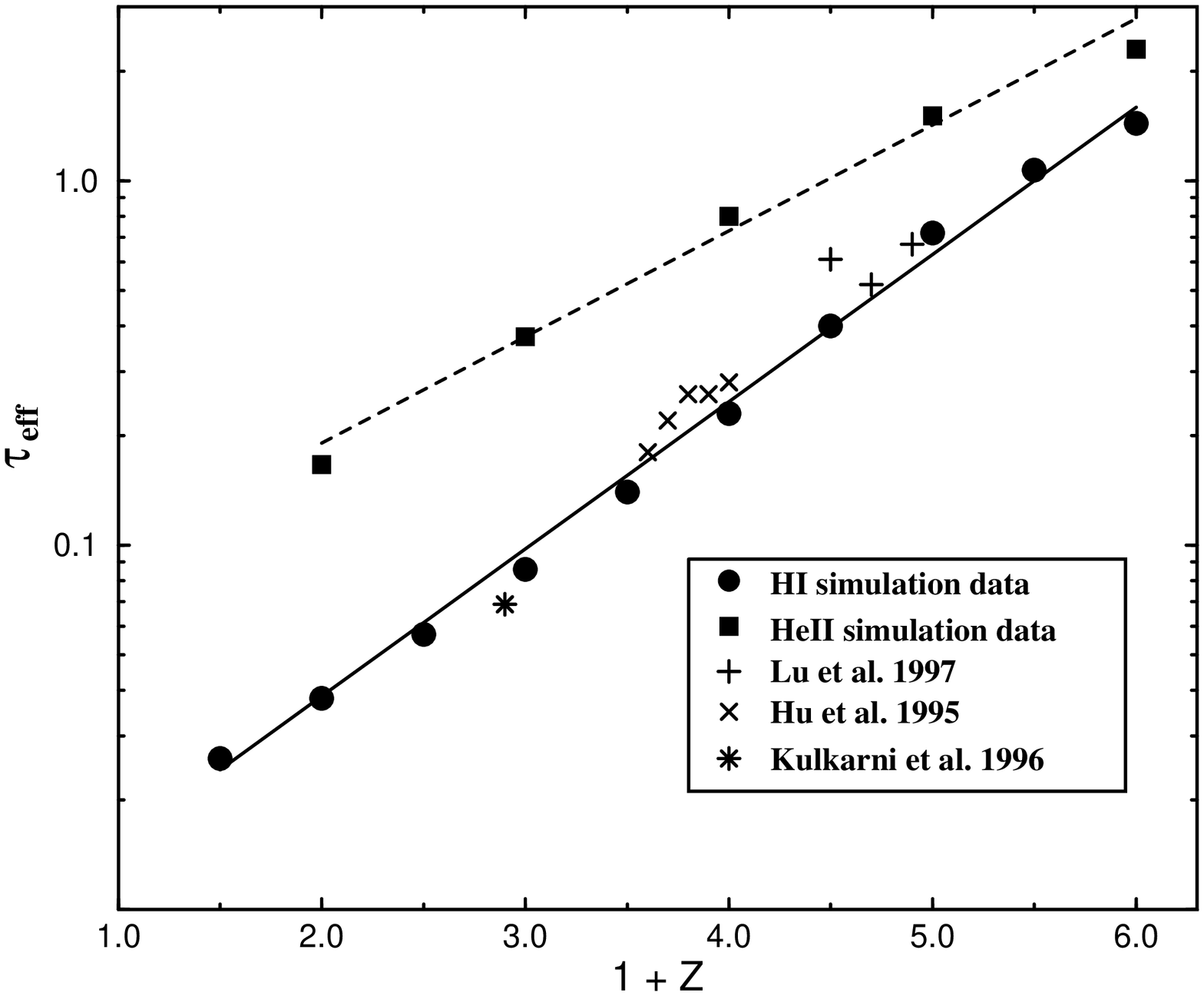}{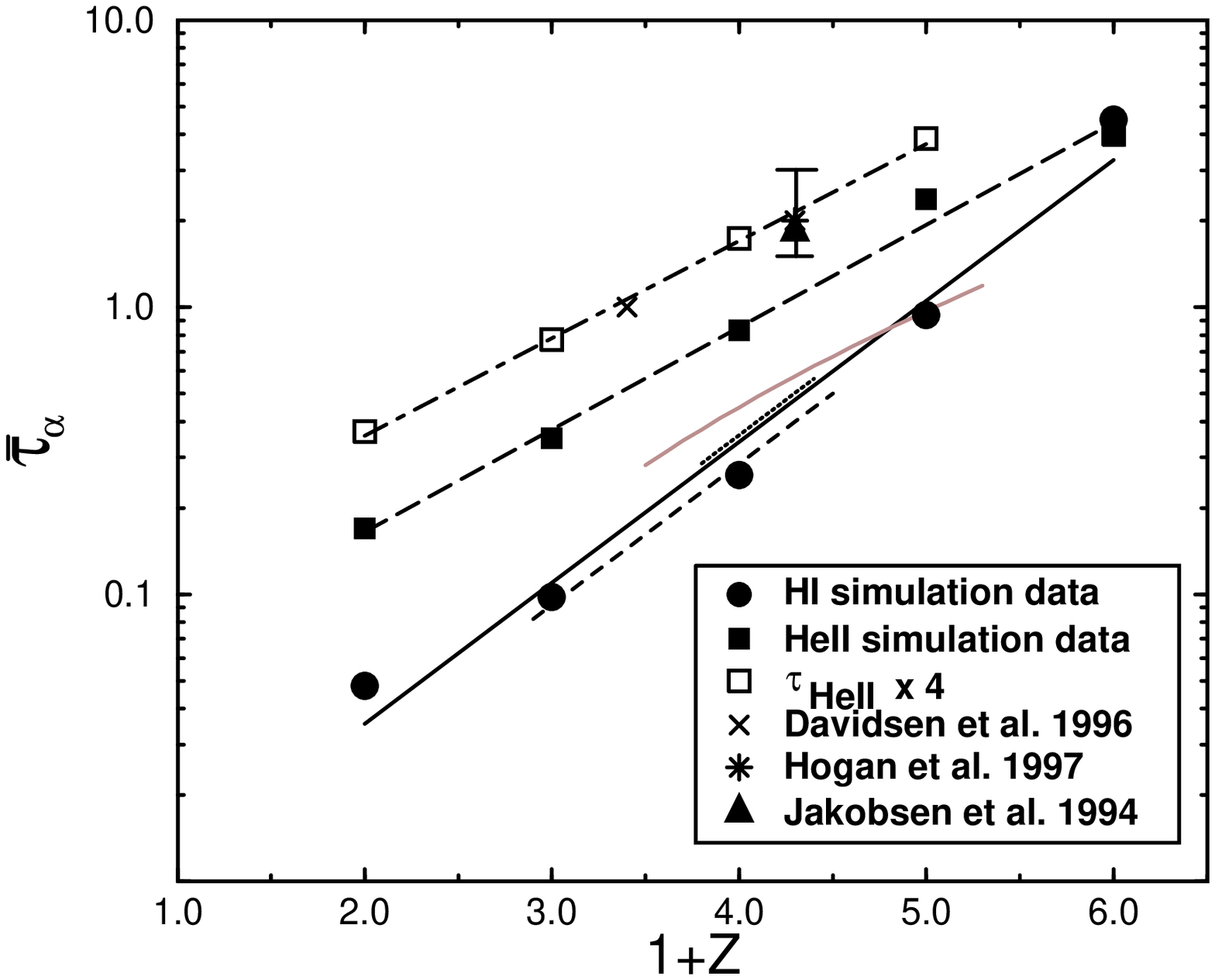}
\caption{Evolution of the effective \Lya opacity from \HI and \HeII line
blanketing, using eq.~(15) of the text and counting only lines
with line center opacity exceeding 0.05.
The evolution is well fit by exponentials of the form
$\tau_0 e^{\alpha(1+z)}$ where $(\tau_0,~\alpha)$ are
($6\times 10^{-3}$, 0.93) and ($5\times 10^{-2}$, 0.67) for
\HI and \HeII respectively. Also shown are the measured effective opacities
using the data taken from the Keck HIRES at $z\sim4$ (Lu \etal 1997),
$z\sim 3$ (Hu \etal 1995), and KPNO at $z\sim 2$ (Kulkarni \etal 1996).}
\label{fig:mod_teff}
\caption{Evolution of the average \Lya opacity, $\bar\tau_\alpha\equiv
-\log\langle e^{-\tau}\rangle$. The values for \HI are shown by the filled
circles. The \HeII opacities are shown by the filled squares. The open squares
show the resulting average opacity after multiplying the \HeII opacities
in the synthetic spectra by a factor of 4. The fit for \HI from Press \etal
(1993) is shown by the gray solid line. The fit to the absorption
measurements of Steidel \& Sargent (1987) is shown by the dotted line, and
to those of Zuo \& Lu (1993) by the
short--dashed line (see text). The \HeII data are taken from
Davidsen \etal (1996), Hogan \etal (1997) (95\%
confidence interval), and the 90\% lower limit of Jakobsen \etal (1994).
The \HI simulation results are fit by $\bar\tau_\alpha^{\rm HI}=
0.0037e^{1.13(1+z)}$ ({\it solid line}). The \HeII data are fit by
$\bar\tau_\alpha^{\rm HeII}=0.032e^{0.82(1+z)}$ ({\it dashed line}),
and $\bar\tau_\alpha^{\rm HeII}=0.075e^{0.78(1+z)}$ ({\it dot--dashed line}).
The average \HeII opacity at $z=5$ for the factor of 4 increase in the
spectral opacities coincides nearly with the solid square at $z=5$, and is
not included in the fit. The average opacities agree closely with the
effective opacities for \HI and \HeII shown in Figure 14, for $z\le3$, so that
almost all the opacity arises from identified absorption lines.}
\label{fig:mod_tavg}
\end{figure}
Figure \ref{fig:mod_teff} shows the evolution of the effective opacity
contributed by all the \HI \Lya absorption lines (for $F_t=0.95$).
The evolution can be fit
to an exponential law $\tau_{\rm eff} \sim \tau_1 e^{\alpha (1+z)}$,
with $\tau_1 = 6 \times 10^{-3}$ and $\alpha = 0.93$.
(The exponential fit is much better than a power--law fit here, as it
better approximates the behavior of the radiation field.)
We also compute the effective opacity from the published line lists of the
four QSOs from Keck HIRES observations at $z\sim3$ (Hu \etal 1995),
the HIRES observation at $z\sim4$ (Lu \etal 1997), and
the KPNO observation at $z\sim2$ (Kulkarni \etal 1996), using
eq. (\ref{eqn:mod_teff}), and counting
only those lines with line center opacity exceeding 0.05.
The results, plotted in Figure \ref{fig:mod_teff}, agree with
our numerical data quite well. To assess the contribution of incompleteness
in the line lists, we have recomputed the effective opacities for the
Hu et al. data by weighting each line by the inverse of the incompleteness,
as given in their Table 3. The resulting effective opacities increase by at
most an additional amount of 0.01--0.02.

Because current measurements of the \HeII \Lya forest are not able to
resolve individual lines, we cannot compare our number counts for the
\HeII \Lya absorbers directly with observations. Higher resolution and
more sensitive instrumentation is required. The situation will improve
with the installation of the Space Telescope Imaging Spectrograph
(STIS) on HST, but at best only the broadest \HeII absorbers will be
resolvable in QSO spectra. We may, however, make an indirect
comparison with the data using the effective opacity. The effective
opacity for \HeII from our simulations is shown in Figure
\ref{fig:mod_teff}. The opacity is well fit by an exponential with
parameters $\tau_1 = 5\times 10^{-2}$ and $\alpha=0.67$.

\subsection{Average opacity}
\label{subsec:avgtau}

\subsubsection{Simulation Results}
\label{subsec:avgtau_sim}

In principle, a contribution to the over-all absorption may arise from
a component of the IGM that is not readily identified as discrete absorbers.
An alternative description of the opacity including all contributions is
given by $\bar\tau_\alpha\equiv\log Q_\alpha^{-1}$, where
$Q_\alpha$ is the observed transmission counting \Lya absorption only
(Press, Rybicki, \& Schneider 1993). From an analysis of 29 low resolution
spectra from the Schneider--Schmidt--Gunn (SSG) QSO survey,
Press \etal derive $\tau_\alpha\approx0.0037(1+z)^{3.46}$ over the
redshift range $2.5<z<4.3$.
In Figure \ref{fig:mod_tavg}, we show the average opacity derived from
our spectra, both for \ion{H}{1} and \ion{He}{2}. The result of Press \etal
for \ion{H}{1} \Lya is shown as a gray solid line.

A comparison with Figure \ref{fig:mod_teff} for the effective
opacity due to identified lines alone shows that very little residual
opacity is found to arise in the simulation that cannot be fully
accounted for by the lines. For $1\le z\le3$, we find
$\bar\tau_\alpha-\tau_{\rm eff}<0.03$. The small differences may be due to
the difficulty of identifying lines with line center opacity smaller than
0.05. (Only lines with line center opacities exceeding
0.05 were included in our estimate of $\tau_{\rm eff}$.) At higher
redshifts the difference increases. It is difficult to assess at this
point, however, whether the additional absorption is due to gas that
has not condensed into absorbing clouds or to incompleteness in the
line finding as a result of line--blending at these high redshifts.
By comparison,
the Press \etal estimate for $\bar\tau_\alpha$ at $z=3$ is 0.45,
exceeding the simulation results by 70\%. The direct line count of the
Hu et al. (1995) data (shown in Figure \ref{fig:mod_teff}), gives
$\tau_{\rm eff}=0.28$. This resurrects an old problem
identified by Jenkins \& Ostriker (1991):\ the
opacities from the SSG QSO spectra greatly exceed those due to
line blanketing based on direct line counting from higher resolution
spectra. Here we point out that the discrepancy persists for line
lists based on the much higher resolution, higher signal-to-noise
spectra from the Keck HIRES. Recomputing $\bar\tau_\alpha$ for the
Keck data would substantially help to clarify the origin, or reality, of
this discrepancy. One possibility is that the difference is an artifact
of the low resolution (25\AA) of the SSG spectra, which could result in an
overestimate of the continuum level by failing to resolve emission features
longward of the QSO \Lya emission line (Steidel \& Sargent 1987). The higher
resolution data (6\AA) of Steidel \& Sargent (1987) result in systematically
lower opacities than derived from the SSG data. We return to this point in
the following section.

There are two QSO lines--of--sight for which measurements of intergalactic
\HeII absorption have been published. At $z\approx3.3$, Jakobsen \etal (1994)
obtain $\tau>1.7$ (90\% confidence), while, for the same QSO sightline,
Hogan \etal (1997) find
$1.5<\tau<3$ (95\% confidence). For a second line--of--sight, Davidsen \etal
(1996) obtain $\tau=1.0\pm0.07$ ($1\sigma$), at $\langle z\rangle\approx2.4$.
We show these limits in Figure \ref{fig:mod_tavg}, along with
$\bar\tau_\alpha$ for \HeII computed from the simulation. The measured values
substantially exceed those of the simulation. The difference may be accounted
for by increasing the size of the break at the \HeII Lyman edge, but would
require a decrease by a factor of $\sim4$ of the estimate for the \HeII
ionization rate from Haardt \& Madau (1996) (shown by the open boxes).
While this would be consistent with the break that would result from a
soft QSO spectrum (Madau \& Meiksin 1994), it is fairly large, and would
require a steep intrinsic QSO spectrum, with $\alpha_Q\approx1.8-2$.
Alternatively, $\Omega_b$ may be increased, and additional sources of \HI
ionizing photons postulated to prevent too large a value for $\tau_{\rm eff}$
due to line--blanketing from the neutral hydrogen. This would require a
doubling of $\Omega_b$ to 0.12, a high value but still in keeping with
the Big Bang nucleosynthesis constraints imposed by measurements of deuterium
in high redshift absorption systems (Tytler \etal 1996). The agreement of the
\HI simulation results with the observed redshift trend of $\tau_{\rm eff}$ in
Figure \ref{fig:mod_teff}, however, would then in part be fortuitous.

\subsubsection{Comparison with TreeSPH}
\label{subsec:avgtau_sph}

We compare our opacity results with those of Croft \etal (1997), based on
a similar SCDM simulation, but using TreeSPH. We summarize the comparison
in Table 5. For $\Omega_b=0.06$ and the
Haardt \& Madau spectrum, the ionization bias for \HI in our simulation is
$b_{\rm ion}^{\rm HI} = 0.0035$ at $z=2$ and $b_{\rm ion}^{\rm HI} = 0.0038$
at $z=3$. For the parameters chosen by Croft \etal for their dynamical run
($\Omega_b=0.05$, half the Haardt \& Madau rates),
$b_{\rm ion}^{\rm HI} = 0.0052$ at $z=2$ and 3. They find for their run
$\bar\tau_\alpha^{\rm HI}=0.16$ at $z=2$, and 0.41 at $z=3$ (from their
Figure 3b). Scaling the opacities in our spectra to
$b_{\rm ion}^{\rm HI}=0.0052$ (and assuming no temperature correction), we find
$\bar\tau_\alpha^{\rm HI}=0.13$ at $z=2$ and 0.32 at $z=3$. 
We thus find only fair agreement with Croft \etal:\ their values exceed ours
by 25--30\%. For \HeII, our simulation ran with $b_{\rm ion}^{\rm HeII}=0.31$
at $z=2.4$ and 0.42 at $z=3.3$. The simulation of Croft \etal ran with
$b_{\rm ion}^{\rm HeII}=0.47$ at $z=2.4$ and $b_{\rm ion}^{\rm HeII}=0.59$
at $z=3.3$. They find $\bar\tau_\alpha^{\rm HeII}=0.73$ at
$z=2.4$ and $\bar\tau_\alpha^{\rm HeII}=2.0$ at $z=3.3$. For the same values of
the ionization bias (and no temperature correction), we obtain
$\bar\tau_\alpha^{\rm HeII}=0.63$ at $z=2.4$ and 1.3 at $z=3.3$. We thus find
a discrepancy in the \HeII opacities as well:\ the values in Croft \etal
exceed ours by 15--50\%.

It is possible that much of the discrepancy is due to differences in
the gas temperatures. Such a difference may arise as a consequence of
the different ionization histories adopted by the two simulations,
particularly in the underdense regions, where most of the \HeII
opacity is generated. An increase in temperature lowers the radiative
recombination rates to \HI and \HeII, and so reduces their respective
column densities for the same total gas density. Both simulations,
however, produce comparable temperatures, 7000--15000 K, over the
range of densities dominating the \HI opacity (overdensities of a
few). Still, a 50\% change in temperature will alter the column
densities by 30\%, so temperature differences may play some role in
accounting for the different opacities. A second possibility is that
the higher Doppler parameters found in the TreeSPH simulation compared
to ours (\S \ref{subsec:doppler}), reflects a true underlying difference
between the simulations. The larger $b-$values would broaden the
saturated lines, increasing their contribution to the average opacity
in direct proportion to the ratio of Doppler parameters. The mean SPH
$b-$value is $\sim30\%$ larger than ours, so that the difference in
the Doppler parameter distributions may account for a large part of the
discrepancy in the opacities.

There is a significant
difference in temperatures in the underdense regions. The TreeSPH
simulation finds $T\lsim6000$ K (Katz \etal 1996).
Our simulation shows a spread in temperatures over approximately
$6000-8000$ K for the underdense gas at $z=3$ (Zhang \etal 1997).
The temperature in the underdense regions is particularly susceptible to
details of the ionization history because the density is too low for
the gas to maintain the temperature at the photoionization thermal
equilibrium value. Allowing for the increase in the radiative
recombination rate on changing the temperature from 8000 K to 6000 K,
the ``temperature corrected'' average opacity values from our
simulation are $\bar\tau_\alpha^{\rm HeII} = 0.71$ at $z=2.4$ and 1.5 at
$z=3.3$. The value at $z=2.4$ now compares very favorably with the
SPH calculation, although a 30\% discrepancy still remains at $z=3.3$.

The temperature dependence of the average opacity actually does not
scale quite so simply, since the absorption profile of the
distribution of the absorbing atoms is sensitive to the temperature as
well. We may assess the effect of a change in temperature
quantitatively when the average opacity is dominated by
line--blanketing, as we in fact find is the case. The opacity is then
given by the average equivalent width of the lines (cf., eq.
[\ref{eqn:mod_teff}]). For optically thin clouds, the equivalent
width scales linearly with the column density, hence with the
radiative recombination rate (the linear part of the
curve--of--growth). For saturated absorption, however, the equivalent
width scales in direct proportion to the Doppler parameter, and
becomes nearly insensitive to the column density and recombination
rate (Spitzer 1978). Since the line--blanketing opacity is dominated
by saturated lines, the temperature sensitivity will be suppressed.
Indeed, depending on the distribution of line center opacities in the
forest, an increase in gas temperature could even result in an {\it
increase} in the average opacity. We may derive the temperature
dependence from eq. (\ref{eqn:mod_teff}), where we substitute a
distribution over line center opacity $\tau_0$ for the distribution
over rest equivalent width $W_0$. Assuming a power law opacity
distribution with slope $\beta$, and using eqs. (\ref{eqn:spec_b}) --
(\ref{eqn:spec_col}), we obtain (assuming the integration range may be
extended from 0 to $\infty$),
\begin{equation}
\tau_{\rm eff}\propto\alpha_{\rm R}^{\beta-1}b^{2-\beta},
\end{equation}
where $\alpha_{\rm R}$ is the radiative recombination coefficient to
either \HI or \HeII. Over the temperature range of interest,
$\alpha_{\rm R}\propto T^{-0.7}$.  For thermally broadened lines,
$b\propto T^{1/2}$. We then find for \HI, with $\beta=1.64$,
$\tau_{\rm eff}\propto T^{-0.27}$. For \HeII, with $\beta=1.73$, we
obtain $\tau_{\rm eff}\propto T^{-0.38}$. We find that this relation
results nearly exactly in the temperature scaling for the \HeII
average opacity found above.

Although an opacity difference of 30\% may not seem large, because
of the weak dependence of opacity on $b_{\rm ion}$ it translates into a
substantial difference in the requirements for the background ionization
field. Normalized to the same value of $\Omega_b$, we find we need to reduce
the Haardt \& Madau estimate of the \HeII ionizing background by an additional
factor of 1.5--2 above the requirements of Croft et al. in order to match to
the \HeII observations. It is critical to resolve
the formation of structures on small scales to ensure convergence to the
correct opacity. The clumping of gas into clouds that produce saturated or
nearly saturated absorption features opens up regions in the spectrum of low
optical depth, and so increases the average transparency. Since most of the
\HeII absorption arises in structures that are underdense, we believe it likely
that the difference in resolution between the two simulations is in part
responsible for the discrepancy in opacities. While our resolution at $z=3$ is
18.75 kpc on the top grid, the resolution in the lowest density regions is
$\sim200$ kpc in the SPH simulation (Hernquist \etal 1996). We provide a more
extensive discussion of the distribution of opacity in Zhang \etal (1997).

\subsection{Flux decrement}
\label{subsec:decrement}

A particularly useful wavelength range to measure the flux decrement
due to the forest is between \Lya and \Lyb in the restframe of the QSO.
This region avoids \Lyb absorption by the forest, enabling an estimate
for the decrement based soley on \Lya absorption. Oke \& Korycansky
(1982) define the flux decrement $D_{\rm A}$ as
\begin{equation}
  D_{\rm A} = \frac{32}{5}\frac{1}{1+z_Q}\int_{(27/32)(1+z_Q)-1}^{z_Q}
        \left[1-\exp(-\tau_\nu)\right]dz , \label{eq:DA}
\end{equation}
where $z_Q$ is the redshift of the QSO. (In practice, an upper wavelength
somewhat short of \Lya is chosen to avoid the broad wing of the QSO \Lya
emission line. Here, we retain the full wavelength range. The difference
should be small.)
\begin{figure}
\plottwo{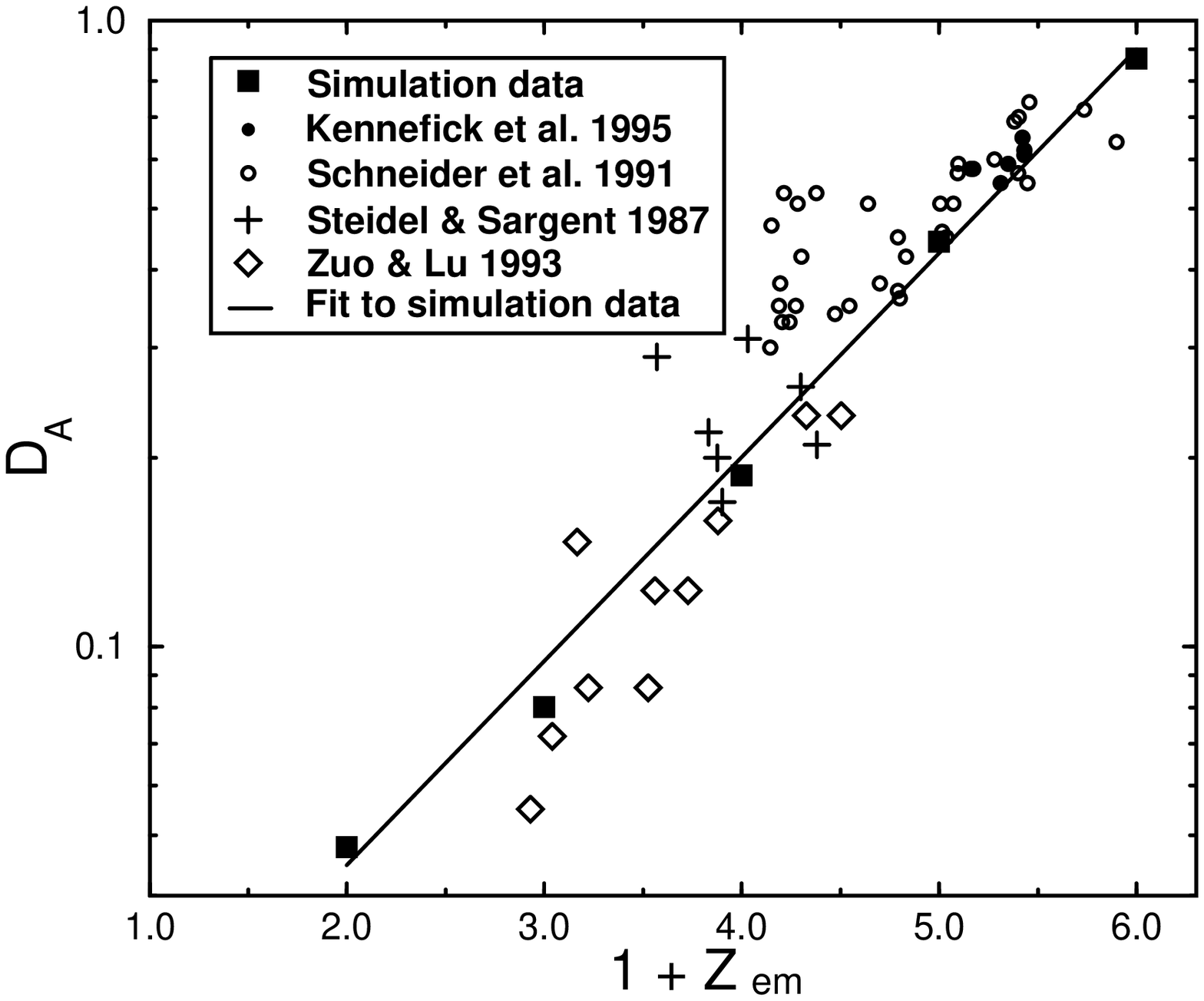}{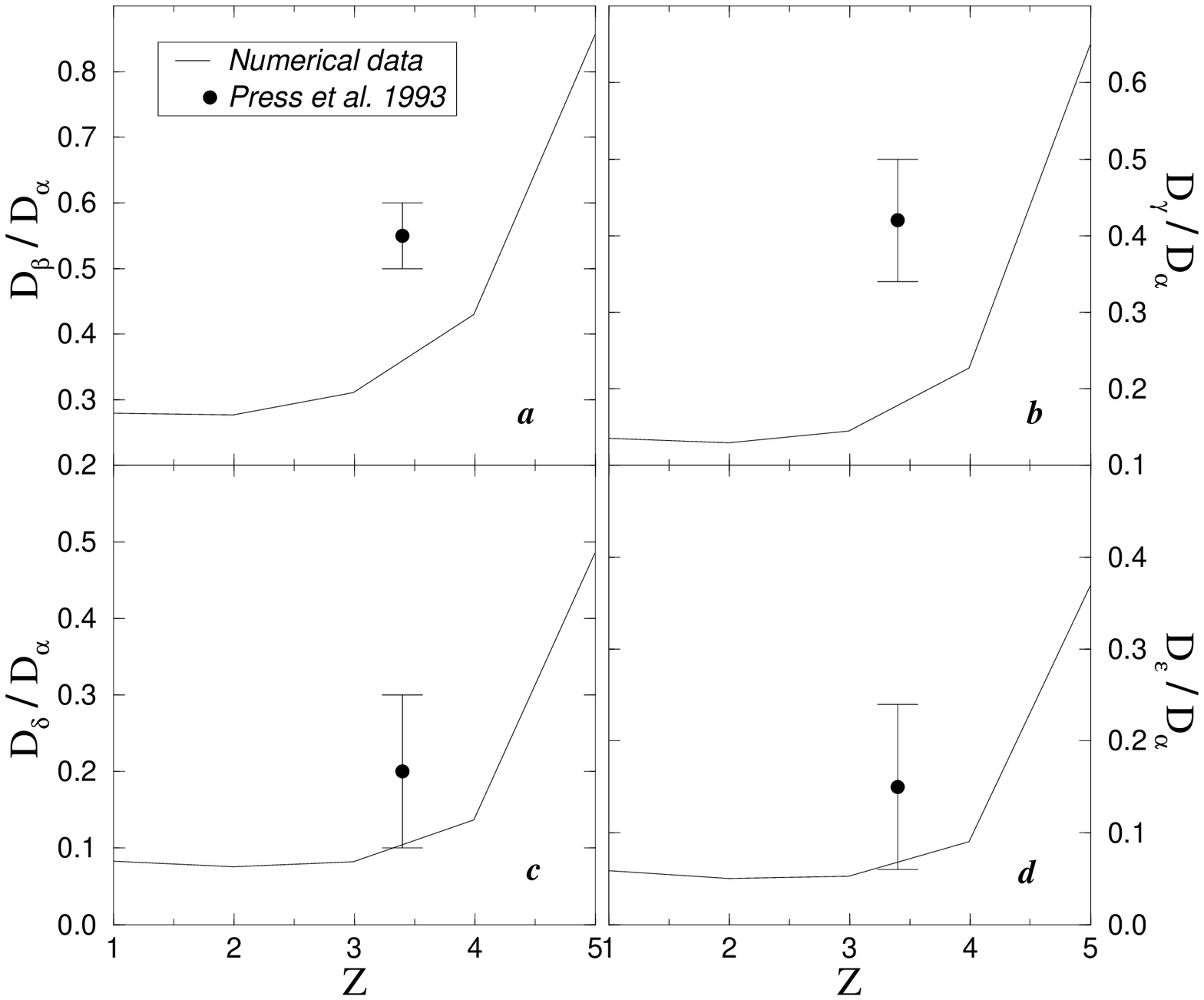}
\caption{Evolution of the \HI flux depression $D_{\rm A}$ between Ly$\alpha$
and Ly$\beta$, as a function of the emission redshift of the QSO.
Also shown are four separate observations and an exponential fit
to the numerical data, $D_{\rm A} = 0.01 e^{0.75(1+z)}$.}
\label{fig:mod_DA}
\caption{Evolution of the average \HI flux depression ratios for the
first several Lyman series transitions:
a) $D_\beta    / D_\alpha$,
b) $D_\gamma   / D_\alpha$,
c) $D_\delta   / D_\alpha$ and
d) $D_\epsilon / D_\alpha$.
Also shown with filled circles and error bars are the values
determined from the Press, Rybicki, \& Schneider (1993) analysis of
the Schneider, Schmidt, \& Gunn (1991) data at $z\sim 3.4$. Their derived
ratios of effective opacities translate to $D_\beta/D_\alpha=0.55\pm0.05$,
$D_\gamma/D_\alpha=0.42\pm0.08$, $D_\delta/D_\alpha=0.20\pm0.10$, and
$D_\epsilon/D_\alpha=0.15\pm0.09$.}
\label{fig:series}
\end{figure}
Figure \ref{fig:mod_DA} plots the evolution of
the \HI $D_{\rm A}$ from $z=5$ to $z=1$ along with an exponential fit to
the data of the form $D_{\rm A}={D_{\rm A}}^0e^{\alpha (1+z)}$, with
$D_{\rm A}^0 = 0.01$ and $\alpha=0.75$.
Also shown are four different measurements of the $D_{\rm A}$ decrement from
Steidel \etal (1987), Schneider \etal (1991), Zuo \& Lu (1993),
and Kennefick \etal (1995). Despite the rather large scatter in the
observational data, the agreement with the simulation results is fairly good,
although the measured values of Schneider \etal are systematically higher than
the numerical data. However, these data are predominantly derived from low
resolution spectra (25\AA), and the improved continuum accuracy permitted
by higher resolution observations can lower their values, as discussed by
Steidel \& Sargent (1987), who obtained 6\AA~
resolution data. The values reported
by Zuo \& Lu were estimated by performing new continuum fits to
the $\sim1$\AA~
resolution data of Sargent \etal (1988). (We use only the estimates from their
Table 1, based on a direct integration of the digital spectra.)
In any event, the numerical results (for both the
column density distribution and $D_{\rm A}$ evolution), may be made to match
the observations more precisely by adjusting the physical parameters of our
simulations, either by changing the radiation intensity or the baryon
density. In particular, the more rapid fall off in the values measured by
Zuo \& Lu toward lower redshifts than given by the simulation would be
consistent with the onset of additional sources of \HI ionizing radiation
at $z<3$, such as young massive stars in forming galaxies.

We may use the $D_{\rm A}$ estimates of Steidel \& Sargent (1987) and Zuo \& Lu
(1993) to estimate the mean intergalactic opacity $\tau_\alpha$ as a function
of redshift. We do so by setting $\tau_\nu=\tau_\alpha(z)$, where
$\nu = \nu_\alpha/ (1+z)$, in eq.(\ref{eq:DA}). For the Steidel \& Sargent
data, we attempted to fit their 7 measurements (we exclude the measurement
of the BAL QSO), by the relations $\tau_\alpha=A(1+z)^{3.46}$ and
$\tau_\alpha=Ae^{\beta(1+z)}$, and minimizing $\chi^2$ using their
estimate of $0.05$ for the error in $D_{\rm A}$. We found neither form was
able to give an acceptable fit. The reason was due to an anomalously large
$D_{\rm A}$ value at $z=2.57$. Removing this point, we obtain $\tau_\alpha=0.0028
(1+z)^{3.46}$ and, fixing $\beta=1.13$ to match the simulation fit,
$\tau_\alpha=0.0039e^{1.13(1+z)}$, both with acceptable values for $\chi^2$.
(Varying the redshift evolution rates did not much improve $\chi^2$.)
For the 10 $D_{\rm A}$ measurements in Table 1 of Zuo \& Lu, we obtain
$\tau_\alpha=0.0022(1+z)^{3.46}$ and $\tau_\alpha=0.0031e^{1.13(1+z)}$, both
with acceptable $\chi^2$ values. We show the fits in Figure \ref{fig:mod_tavg}.

Press, Rybicki, \& Schneider (1993) compute the average
flux decrements of the first several Lyman series transitions
(Ly$\beta$, Ly$\gamma$, Ly$\delta$ and Ly$\epsilon$) relative to
the Ly$\alpha$ average flux decrement. They express their results in
terms of ratios of the mean equivalent widths (normalized by the rest
wavelength of the respective transition), or equivalently, in
terms of the ratios of the mean opacities $\bar\tau_\alpha$,
$\bar\tau_\beta$, etc.
These ratios are computed from our simulation by scaling the line center
opacities by the appropriate oscillator strengths and wavelengths. Each
average opacity is meant to include the absorption contribution due to
a single component of the Lyman series.
In analogy to $D_{\rm A}$, we may define $D_i=1-e^{-\bar\tau_i}$ for
the $i^{th}$ term of the Lyman series. Hence, $D_\alpha=D_{\rm A}$.
The results are plotted as a function of redshift
in Figure \ref{fig:series}. Also shown are the corresponding
calculations by Press, Rybicki, \& Schneider (1993) for the observed
data of Schneider, Schmidt, \& Gunn (1991) at $z\sim 3.4$, reexpressed
in terms of the flux decrements $D_i$. The observed values are higher than
our predicted numbers, especially for the Ly$\beta$ and Ly$\gamma$ transitions.
For a pure power--law column density distribution for the \Lya forest,
it is straightforward to show that the ratios of $\bar\tau$'s
(when convergent), are given by $\bar\tau_i/\bar\tau_\alpha=
(\lambda_i f_i/\lambda_\alpha f_\alpha)^{\beta-1}$.
Thus, the difference is likely a reflection of the steeper column density
distribution found in the simulations compared to the rather flat
distribution inferred from the Schneider \etal data:\ Press \& Rybicki (1993)
obtain $\beta=1.43\pm0.04$. It is noteworthy that our results for these
ratios agree well with those of Miralda--Escud\'e \etal (1996), especially
considering that the cosmological models and adopted radiation fields are
different in the two simulations.

\section{Summary}
\label{sec:stat_sum}

We have developed a spectral synthesis and analysis procedure, based
on fitting the absorption features to Voigt profiles. The procedure is capable
of resolving the lowest (as well as the higher) column density
absorption features in numerical simulations.  We have applied our
procedures to several numerical simulations of the \Lya forest in a
standard CDM model with a self--consistent treatment of the dark matter,
the baryonic matter, a chemical reaction network of hydrogen and helium
components, radiative cooling, and the Haardt \& Madau (1996) ionizing
radiation field based on QSOs as the dominant sources of the radiation.
With the adequate modeling of microphysical and
radiation processes, we are able to reproduce the essential
observational data down to the completeness \HI column densities of
our simulations, approximately $10^{12}$ \cm2 at redshift $z=3$.  The
advances in our line analysis methods are essential in resolving the
low density, optically thin \HI \Lya features, especially for comparing
results to recent high resolution observed data such as taken by the
Keck HIRES. We find excellent agreement with the observed data for our
adopted simulation parameters, including the slopes and shapes of the column
density and equivalent width distributions, the value and distribution
of Doppler parameters, and the evolutionary histories of the line
number, flux decrement, and opacities. We comment that we cannot reproduce
both the high \HI mean opacity estimate of Press \etal (1993) and the \HI
effective opacity due to line--blanketing. The Press \etal opacity
substantially exceeds the opacity due to line--blanketing as determined
directly from the line lists obtained using the Keck HIRES. We find instead
that the average opacity in the simulation agrees very closely to the
line--blanketing opacity for $1\le z\le 3$, with
$\bar\tau_\alpha-\tau_{\rm eff}<0.03$ over this redshift range. It
is possible that the low resolution of the spectra analyzed by Press \etal
resulted in an overestimate of the average intergalactic \HI opacity. We find
good agreement with the values of the average flux decrement $D_{\rm A}$
measured by Steidel \& Sargent (1987), Zuo \& Lu (1993), and
Kennefick \etal (1995), over the redshift range $2\lsim z\lsim5$, as well as
with the mean intergalactic \Lya opacity we infer from these values.
Since the opacity is sensitive to the intensity of the UV photoionizing
background, the agreement over the wide redshift range is noteworthy, and
supports QSOs as the dominant sources of the ionizing radiation field.

The evolution in the \HI line number density we obtain is somewhat too rapid
compared to observations at the highest redshifts ($z>3.5$). The rate of
evolution, however, is sensitive to the assumed
evolution of the radiation field, which becomes increasingly uncertain
at these redshifts. We are able to match the normalization in the
number counts of the \Lya forest systems at the low column density end
($\NHI\lsim14$ \cm2) assuming the Haardt \& Madau QSO-dominated UV
radiation background and a baryon density consistent with standard
nucleosynthesis limits, but curvature in the distribution results in
a significant deficit of lines at the high end
($\NHI>{\rm few\,}\times10^{16}$ \cm2). It is possible, however,
that the discrepancy is due to the finite box size and resolution of
the simulation (and the neglect of self--shielding effects for the very high
column densities), rather than an intrinsic consequence of SCDM.

We have explored the effects of grid resolution by performing
analogous statistical calculations of the \Lya absorbers on higher
resolution grids and found that our results are not sensitive to cell
size at the low column density end, down to the completeness column
density limit. We find that (for our spatial grid resolutions)
intrinsic line blending is the dominant cause of incompleteness in the
generated line lists at high redshifts. The grid resolution (or, in
the case of SPH calculations, the effective resolution defined by
particle counts in the void regions) can be important in setting the
completeness density at smaller redshifts.

We compare our results with those of a similar simulation with nearly
identical cosmological parameters and power--spectrum performed by
Hernquist \etal (1996) using TreeSPH, as re-analyzed by Dav\'e \etal (1997)
using Voigt profiles and by Croft \etal (1997) for the opacity distributions.
While there is generally good agreement between the two
simulation results, we find that a power--law \HI column density distribution
extends to substantially lower column densities than they find. We also
find somewhat lower Doppler parameters and a narrower Doppler parameter
distribution, in agreement with Keck HIRES measurements.
Because Dav\'e \etal introduce observational systematics into their spectra
in order to mimic the measured spectra, while we work directly
with the opacity assuming no observational systematics, much of the
discrepancy is likely due to the differences in the analysis procedures.
We nonetheless do find some differences in the analysis--independent
average opacities, as given by Croft et al. The TreeSPH average \HI opacities
exceed our values by 25--30\%, and the \HeII opacities by up to 50\%.
The discrepancies can be accounted for in part by the different ionization
histories adopted for the two simulations, and the resulting differences in
gas temperatures and ionization fractions. The differences, however, may also
be due in part to the lower resolution of the TreeSPH calculation.

The \HeII opacities we obtain are substantially smaller than
those measured. We are able to reach agreement with the observations by
reducing the \HeII ionizing background by a factor of 4 relative to
the estimate of Haardt \& Madau (1996). The smaller intensity requires a soft
intrinsic QSO spectrum, with $\alpha_Q\approx1.8-2$.

Although we have not pursued a detailed analysis of cloud morphologies
in this paper, we do note that the geometrical shapes of the clouds
are found to be linked to the local thermal and gravitational
environments (see also Miralda--Escud\'e \etal 1996). Spheroidal
clouds are concentrated at the intersections of the filaments which
commonly form in CDM--like models, and are typically more massive (and
gravity confined) than the elongated clouds which tend to lie along
the filament strands.  Our simulations also indicate that a
significant percentage of low column density absorption lines come
from local density enhancements in the underdense void regions.
Further investigations of cloud morphologies and ``minivoid''
absorption features are presented in a companion paper (Zhang \etal 1997).

\acknowledgements
We are pleased to thank Ed Bertschinger, Renyue Cen, Lars Hernquist,
Jorde Miralda--Escud\'e, Jerry Ostriker, and David Weinberg
for useful conversations, and Martin White for helpful comments.
This work is supported in part by the NSF under the auspices of the
Grand Challenge Cosmology Consortium (GC$^3$). The computations
were performed on the Convex C3880 and the SGI Power Challenge at the
National Center for Supercomputing Applications, and the Cray C90 at
the Pittsburgh Supercomputing Center under grant AST950004P. A.M. thanks
the William Gaertner Fund at the University of Chicago for support.



\clearpage

\begin{table}
\vskip30pt
\begin{tabular}{|c|c|c|c|c|c|c|c|c|c|}
\tableline
Run & $\Omega_0$ & $\Omega_b$ & $h$ & $\sigma_{8h^{-1}}$ &
$L$ (Mpc) & $\Delta x$ (kpc) & $M_b (M_\odot)$ &
$\Delta M_b (M_\odot)$ & $\Delta M_d (M_\odot)$
\\ \hline \hline
9.6--Top  & 1 & 0.06 & 0.5 & 0.7 & 9.6 &
           75    & $3.7\times10^{12}$ & $1.8\times10^6$ & $2.9\times10^7$
\\ \hline
9.6--Sub  & 1 & 0.06 & 0.5 & 0.7 & 9.6 &
           18.75 & $5.8\times10^{10}$ & $2.7\times10^4$ & $4.6\times10^5$
\\ \hline
3.2--Top  & 1 & 0.04 & 0.7 & 1.0 & 3.2 &
           25    & $1.8\times10^{11}$ & $8.5\times10^4$ & $1.7\times10^7$
\\ \hline
3.2--Sub  & 1 & 0.04 & 0.7 & 1.0 & 3.2 &
           6.25  & $2.8\times10^{9}$  & $1.3\times10^3$ & $2.7\times10^5$
\\ \hline
\end{tabular}
\caption{
The physical and computational parameters of the different \Lya forest
simulations.
$\Omega_0$ is the cosmological density parameter,
$\Omega_b$ the baryonic mass fraction,
$h$ the Hubble parameter,
$\sigma_{8h^{-1}}$ the fluctuation normalization in a sphere of
radius $8h^{-1}$ Mpc,
$L$ the comoving box size,
$\Delta x$ the comoving cell size,
$M_b$ the total baryonic mass in the simulation,
$\Delta M_b$ the initial mass of baryons in a single cell and
$\Delta M_d$ the dark matter particle mass.
The ``Top'' and ``Sub'' in the run labels refer to the top or sub grid
calculations. We use $128^3$ cells for both the top and sub grids,
and a refinement factor of 4, hence the effective grid resolution
of the sub grids is $512^3$. Results for the small box, high amplitude model
were reported in ZAN95.
}
\label{tab:sim_res}
\end{table}

\begin{table}[htb]
\vskip30pt
\begin{tabular}{|l|c|c|c|c|} \hline
Instrument & $\Delta z$  & $\Delta v$ (\kms) & $\Delta \lambda_{\rm HI}$ (\AA)
 & $\Delta \lambda_{\rm HeII}$ (\AA) \\
\hline \hline
numerical spectrum & $2\times10^{-6}$ & 0.6 & $\sim0.01$ & $\sim0.002$ \\
\hline
KECK/HIRES  & $(1.8\sim2.8)\times10^{-5}$ & $5.3\sim8.4$ & $0.09\sim0.14$ & N/A
\\ \hline
KPNO/KPE    & $6\times10^{-5}$   & 18  & $\sim0.3$ & N/A \\ \hline
ASTRO-2/HUT & $2.5\times10^{-3}$ & 750 & N/A & 3 \\ \hline
HST/FOC     & $1.6\times10^{-2}$ & 5000 & N/A & 20 \\ \hline
\end{tabular}
\caption{
The spectral resolution in our numerical data is shown along with
some relevant observational instruments.
The wavelength resolutions are calculated
in the observer's frame for a typical redshift $z=3$.
$N/A$ is entered in the table when the relevant wavelength range is not covered
by the instrument.
Our spectral resolution is more than adequate to cover the grid cell intervals,
which, at $z=3$, are
$\Delta z = 1\times 10^{-4}$ and $2.5\times 10^{-5}$ for the top and sub
grids respectively.
}
\label{tab:spec_res}
\end{table}

\begin{table}[htb]
\vskip30pt
\begin{tabular}{|c|c|c|c|c||c|c|c|c|} \hline
Line \# &
\multicolumn{4}{c||}{Threshold method} &
\multicolumn{4}{c |}{Deblending method} \\ \hline \hline
  & $v$ (\kms) & $N$ (\cm2) & $W_0$ (\AA) & $b$ (\kms) &
    $v$ (\kms) & $N$ (\cm2) & $W_0$ (\AA) & $b$ (\kms)  \\ \hline
1 & 3328.2 & 2.2e14 &  0.09  & 33.8  & 3328.2 & 3.1e14 & 0.14 & 33.7 \\ \hline
2 & 3519.6 & 7.9e12 &  0.04  & 24.2  & 3519.6 & 1.5e13 & 0.07 & 24.2 \\ \hline
3 & 4106.4 & 1.3e13 &  0.06  & 30.3  & 4106.4 & 2.1e13 & 0.10 & 30.3 \\ \hline
  &        &      &        &         & 4301.6 & 4.2e14 & 0.36 & 25.2 \\
4 & 4392.2 & 1.2e15 &  1.15  & 168.5 & 4449.7 & 7.2e14 & 0.64 & 45.6 \\
  &        &      &        &         & 4608.8 & 2.0e13 & 0.09 & 23.5 \\ \hline
5 & 5146.4 & 9.9e13 &  0.32  & 66.4  & 5112.6 & 6.8e13 & 0.21 & 24.6 \\
  &        &      &        &         & 5185.7 & 4.0e13 & 0.17 & 40.3 \\ \hline
\end{tabular}
\caption{
The spectral properties of a few typical lines
(extracted from Figure \protect\ref{fig:spec_com}) as derived from the
threshold
and deblending methods.
Lines \#1--3 are single isolated features; lines \#4 and \#5 are blended
below the transmission cutoff $F_t = 0.7$.
$v$ is the line center position in velocity space,
$N$ is the extracted column density, $W_0$ the equivalent width, and
$b$ the Doppler parameter.
}
\label{tab:spec_com}
\end{table}

\begin{table}[htb]
\vskip30pt
\begin{tabular}{|c|c|c|l|} \hline
$\gamma$ & $W_0$ limit (\AA) & Redshift interval & Observations \\ \hline
\hline
$2.83\pm0.40$ & 0.36 & 1.5 -- 3.5 & $2.75\pm0.29$ (\lu91)  \\ \hline
$5.28\pm0.60$ & 0.32 & 2.5 -- 4.0 & $1.89\pm0.28$ (\bec94) \\ \hline
$5.28\pm0.60$ & 0.32 & 2.5 -- 4.0 & $3.19\pm0.76$ (\hu95; \luu96) \\ \hline
$2.95\pm0.40$ & 0.32 & 1.5 -- 3.5 & $2.31\pm0.40$ (\mur86) \\ \hline
$0.24\pm0.03$ & 0.32 & 0.5 -- 1.5 & $0.60\pm0.62$ (\bah93) \\ \hline
$0.45\pm0.01$ & 0.24 & 0.5 -- 1.5 & $0.42\pm0.42$ (\bahh96) \\ \hline
$4.84\pm0.36$ & 0.16 & 2.5 -- 4.0 & $1.32\pm0.24$ (\bec94) \\ \hline
$4.84\pm0.36$ & 0.16 & 2.5 -- 4.0 & $3.28\pm0.49$ (\hu 95; \luu96) \\ \hline
$2.81\pm0.41$ & 0.14 & 1.5 -- 3.5 & $2.15\pm0.51$ (\gia91) \\ \hline
\end{tabular}
\caption{
A comparison between the exponent $\gamma$ for the line number evolution
in our numerical simulations, and in several different observations.
The exponent is tabulated as a function of the minimum
equivalent width ($W_0$) cutoff, and the redshift interval is chosen
to match the different observations.
}
\label{tab:stat_ew}
\end{table}

\begin{table}[htb]
\vskip30pt
\begin{center}
\begin{tabular}{cccc||cccc} \hline
\multicolumn{4}{c}{\HI Opacity} &
\multicolumn{4}{c}{\HeII Opacity} \\ \hline \hline
$z$	& $b_{\rm ion}$ & $\bar\tau_\alpha^{\rm sim}$  &
$\bar\tau_\alpha^{\rm SPH}$ &
$z$	& $b_{\rm ion}$ & $\bar\tau_\alpha^{\rm sim}$  &
$\bar\tau_\alpha^{\rm SPH}$ \\ \hline
&&&&2.4	&  0.31  &  0.49  &  \nodata \\
2	&  0.0035  &  0.10  &  \nodata &2.4	&  0.47  &  0.63  &  0.73 \\
2	&  0.0052  &  0.13  &  0.16 & 2.4	&  0.47  &  0.71$^*$  & 0.73 \\
&&&&2.4	&  1.2   &  1.1   &  \nodata \\
\hline
&&&&3.3	&  0.42  &  1.1   &  \nodata \\
3	&  0.0038  &  0.26  &  \nodata & 3.3	&  0.59  &  1.3   &  2.0 \\
3	&  0.0052  &  0.32  &  0.41 & 3.3	&  0.59  &  1.5$^*$   &  2.0 \\
&&&&3.3	&  1.7   &  2.2   &  \nodata \\
\end{tabular}
\end{center}
\caption{
A comparison between the average opacity found
in our simulations, $\bar\tau_\alpha^{\rm sim}$, and that found
by Croft et al. (1997), $\bar\tau_\alpha^{\rm SPH}$,
based on a similar SCDM simulation but using TreeSPH. 
The values marked by an asterisk for HeII take into account the
possible effect of the difference in gas temperatures between
the two simulations (see text).
}
\label{tab:avgtau}
\end{table}

\end{document}